\documentclass[epj]{svjour}  
\usepackage{geometry}                		% See geometry.pdf to learn the layout options. There are lots.
\usepackage{graphicx}				% Use pdf, png, jpg, or eps§ with pdflatex; use eps in DVI mode
\usepackage[utf8]{inputenc}									
\usepackage[hidelinks]{hyperref}
\usepackage{textcomp}
\usepackage{amssymb,latexsym,graphicx,subcaption,amsmath,mathtools, hyperref,siunitx,mathrsfs,bm,color,tikz,booktabs,float} % packages used by K. Yamamoto
\DeclareMathOperator{\tr}{tr} % operator used by K. Yamamoto
\usepackage[numbers]{natbib}
\usepackage{graphicx}

\title{Nature's forms are frilly, flexible, and functional}
\author{Kenneth K. Yamamoto \inst{1} \and Toby L. Shearman \inst{2} \and Erik J. Struckmeyer \inst{2} \and John A. Gemmer \inst{3} \and Shankar C. Venkataramani \inst{2}}
\date{\today}

\institute{Department of Mathematics, Southern Methodist University, Dallas TX 75275 \and School of Mathematical Sciences, University of Arizona, Tucson AZ 85721 \and Department of Mathematics and Statistics, Wake Forest University, Winston Salem NC 27109}
\begin{document}

\abstract{
A ubiquitous motif in nature is the self-similar hierarchical buckling of a thin lamina near its margins. This is seen in leaves, flowers, fungi, corals and marine invertebrates. We investigate this morphology from the perspective of non-Euclidean plate theory. We identify a novel type of defect, a branch-point of the normal map, that allows for the generation of such complex wrinkling patterns in thin elastic hyperbolic surfaces, even in the absence of stretching. We argue that branch points are the natural defects in hyperbolic sheets, they carry a topological charge which gives them a degree of robustness, and they can influence the overall morphology of a hyperbolic surface without concentrating elastic energy. We develop a theory for branch points and investigate their role in determining the mechanical response of hyperbolic sheets to weak external forces.  
 % end of PACS codes
} %end of abstract

\maketitle

%%%%%%%%%%%%%%%%%%%%%%%%%%%%%%%%%%%%%%%%%%%%%%%%
%
%		Introduction
%
%%%%%%%%%%%%%%%%%%%%%%%%%%%%%%%%%%%%%%%%%%%%%%%%

\section{Introduction} \label{sec:intro}

{\em ``We have built a world of largely straight lines -- the houses we live in, the skyscrapers we work in and the streets we drive on our daily commutes. Yet outside our boxes, nature teems with frilly, crenellated forms, from the fluted surfaces of lettuces and fungi to the frilled skirts of sea slugs and the gorgeous undulations of corals."} --Margaret Wertheim \cite{wertheim2016corals}.

Leaves, flowers, fins and wings are examples of the ubiquity of frilly, crenellated forms in nature.
\begin{figure}[htbp]
        \begin{subfigure}[t!]{0.35\linewidth}
                \centering
                {\includegraphics[height=2cm, keepaspectratio, trim={0cm, 0, 0cm, 0}, clip]{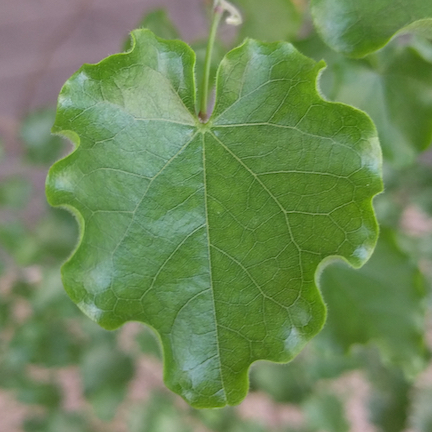}}
                \caption{}
                \label{fig:enr2leaves}
        \end{subfigure}%
        \begin{subfigure}[t!]{0.35\linewidth}
                \centering
                {\includegraphics[height=2cm, keepaspectratio, trim={0cm, 0, 0cm, 0}, clip]{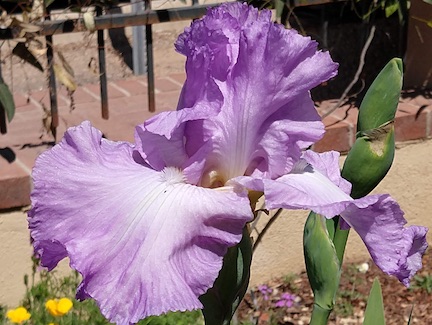}}
                \caption{}
                \label{fig:purpleiris}
         \end{subfigure}%
        \begin{subfigure}[t!]{0.3\linewidth}
                \centering
                {\includegraphics[angle=-90,width=1.5cm]{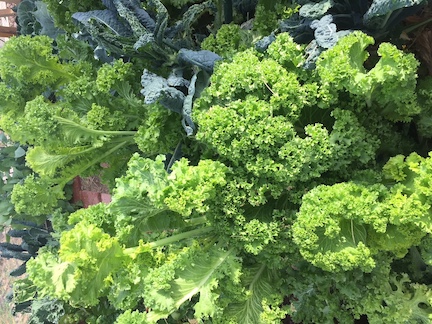}}
                \caption{}
                \label{fig:mustard}    
            \end{subfigure}
            
            \begin{subfigure}[t!]{.35\linewidth}
            \centering
             {\includegraphics[height=1.6cm]{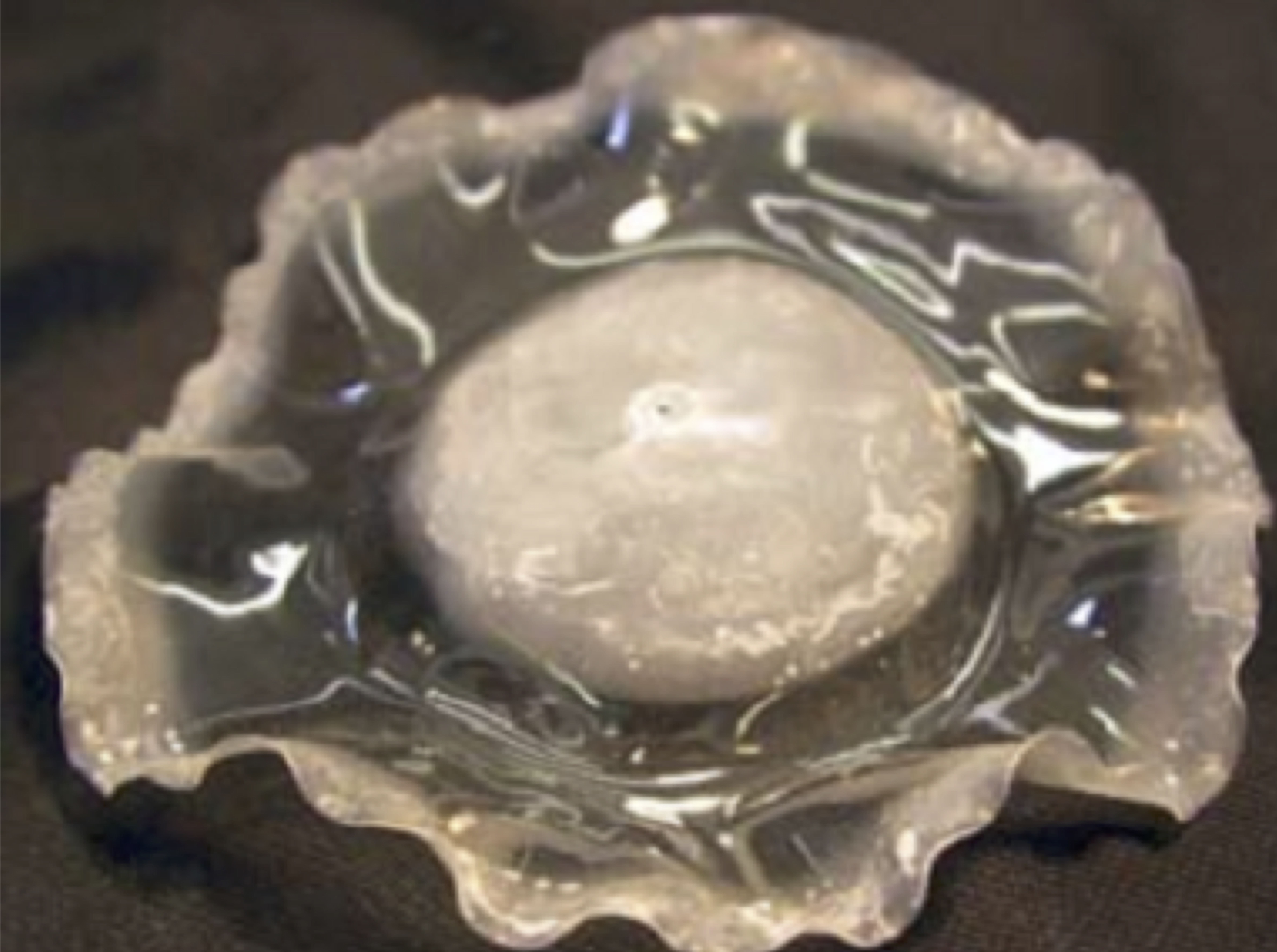}}
                \caption{}
                \label{fig:hydrogel}  
                \end{subfigure}%
                \begin{subfigure}[t!]{.65\linewidth}
                \centering
                {\includegraphics[height=1.6cm]{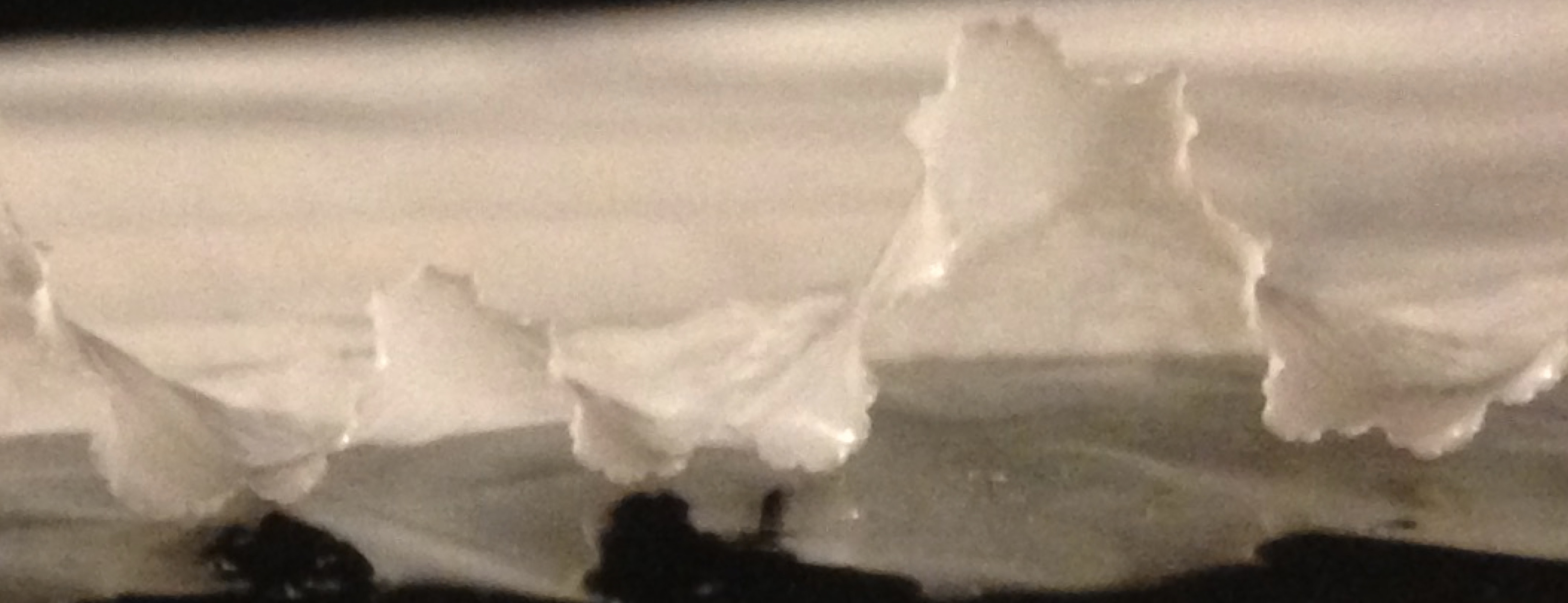}}
                \caption{}
                \label{fig:trasfbag}  
                \end{subfigure}%
        \caption{(a) A leaf with regular undulations (photo by TS). (b) An Iris with 3 generations of undulations (photo by SV). (c) Curly mustard leaves with multiple generations of buckling (photo by J Watkins, U. Arizona). (d) Hydrogel disk (photo by Eran Sharon, Hebrew U. Israel). (e) Torn trash bag (photo by JG).
     }
        \label{fig:examplesofnaturalobjects}
\end{figure}
Fig.~\ref{fig:examplesofnaturalobjects}(a-c) displays some of the complex shapes that result from such hierarchical, ``multi-scale" buckling in leaves and flowers. A relation between these buckling patterns and the growth of a leaf at its margins was first identified by Nechaev and Voituriez \cite{nechaev2001plant}; see also \cite{sharon2002buckling,eran2004leaves,sharon2007geometrically,LiangMaha2011,sharon2018mechanics}. Indeed, a wavy pattern can be induced in a naturally flat leaf by application of the growth hormone auxin to the margins \cite{eran2004leaves} or through genetic mutation \cite{nath2003genetic}. This phenomenon is not restricted to living organisms, where it might be explained as a genetic trait selected through evolution; it is also seen in torn plastic sheets \cite{sharon2002buckling, sharon2007geometrically,audoly2003self} and swelling hydrogels \cite{efrati2007spontaneous, klein2007shaping, kim2012designing}; see Fig.~\ref{fig:examplesofnaturalobjects}(d-e).

The emergence of such patterns results from the sheet deforming to relieve growth induced residual prestrain \cite{goriely2005differential, ben2005growth}. In the non-Euclidean framework of elasticity such strains are encoded in a ``target" Riemannian metric $\mathbf{g}$ which locally measures the preferred distance between between material coordinates \cite{efrati2009elastic, Efrati2013Metric}. Specifically, growth is naturally associated with a hyperbolic metric since the local distance between coordinates is expansive while spherical metrics are associated with atrophy. In this framework, the problem of determining strain-free configurations is equivalent to the mathematical problem of computing isometric immersions of $\mathbf{g}$. Consequently, complex patterns in elastic sheets are known to arise in systems that preclude the existence of isometric immersions by, for example, incompatible boundary conditions \cite{bella2014metric} or confining the sheet to a curved surface \cite{bella2014wrinkles,Tobasco2021Curvature,Tobasco2020Principles}. On the other hand, many growth patterns generate residual in-plane strains with smooth hyperbolic Riemannian metrics on bounded domains which can always be immersed in $\mathbb{R}^3$ by smooth isometries \cite{han2006isometric}. Why then do we observe self-similar buckling patterns in sheets with an intrinsic hyperbolic metric? Why are frilly, crenellated forms ubiquitous in nature and what potential evolutionary benefits arise from these shapes? 

In this paper we investigate these questions by studying the relationship between hyperbolic geometry and the mechanics of thin objects. Building on the work in \cite{EPL_2016}, we investigate and highlight the role that a novel type of topological defect, called a branch point for the normal map \cite{EPL_2016}, plays in not only the selection of patterns but the mechanics of hyperbolic thin sheets. 

\subsection{Multiple scale behaviors in singularly perturbed, energy driven, pattern formation}
\begin{figure}%[tbhp]
\centering
\includegraphics[width=0.95\linewidth]{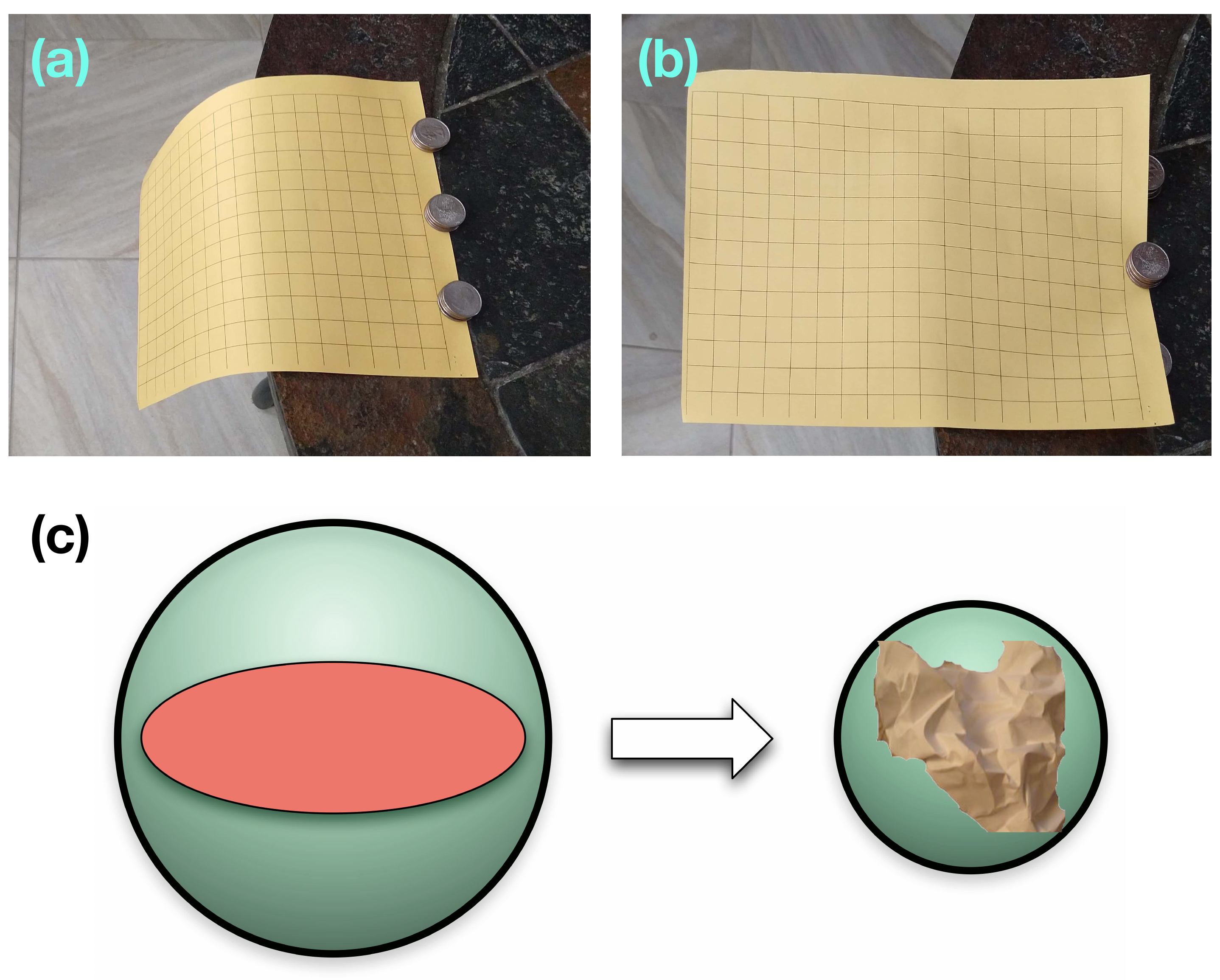}
\caption{(a-b) A letter size paper with a $\frac{1}{2}$ inch grid is held/supported by 3 stacks of 5 quarters, placed 2 inches apart. (a) The sheet is free to bend over the edge of the table without stretching since there is no curvature in the transverse direction. (b) Two of the stacks of quarters are now supporting the sheet from below to introduce a transverse curvature in the sheet. The sheet is now ``rigid" and can no longer bend over the edge without stretching. (c) A thought experiment where a flat sheet is confined within a small sphere causing it to crumple.}
\label{fig:sheet}
\end{figure}

Before beginning a discussion of non-Euclidean elasticity and hyperbolic geometry, we first step back and discuss the larger context of this system, namely energy driven pattern formation in thin elastic sheets. The physics of thin elastic sheets has two key features 
\begin{enumerate}
    \item Multiple energy scales: it is much easier to bend a thin sheet than to stretch it.
    \item Geometric frustration: the fact that stretching and bending are not `independent degrees of freedom'.
\end{enumerate}
Geometric frustration is a consequence of Gauss' Theorema Egregium, which implies that the product of the principal curvatures is an invariant for all deformations that do not stretch the sheet \cite{stoker}. This result is illustrated in Fig.~\ref{fig:sheet}(a-b) in which the equilibrium shape of sheet with an intrinsically flat metric is determined by the interplay between the low cost of bending $E_b$, the high cost of stretching $E_s$, and  geometric rigidity, i.e.  the ``frustration" of stretching energy inhibiting local bending in two independent directions.

An argument for the occurrence of singularities/microstructure in systems with multiple energy scales goes as follows: 
\begin{enumerate}
\item The weak energy scale involves higher order derivatives than the strong energy scale. Specifically, bending involves the curvature while stretching involves only the deformation gradient.
\item Computing the variational derivative results in a singularly perturbed system of Euler-Lagrange equations. 
\item Solutions to singular perturbation problems can contain boundary layers and small scale structures. 
\end{enumerate}
This chain of arguments applies, for example, to ``explain" the microstructure that emerges when crumpling of intrinsically flat thin sheets and is as follows. Gauss' theorem implies that an unstretched sheet with a flat metric has one `locally straight' direction at every point which precludes the confinement of a thin elastic sheet within a sufficiently small sphere without stretching \cite{immersion_thm}. A uniformly stretched configuration will be energetically prohibitive and thus the sheet will minimize its stretching energy $E_s$ by adopting a non-uniform, multi-scale, crumpled configuration, even when the compression is done in a uniform, large-scale manner; see Fig. \ref{fig:sheet}(c). In particular, the energy $E_b + E_s$ in a crumpled sheet condenses on to a network of ridges \cite{science.paper} that meet at point-like vertices \cite{benAmar1997Crumpled}, and outside these defects, the sheet is essentially stress-free \cite{witten2007stress}. 

A refinement of the above ideas is to argue that the small scale structures are determined by a balance between the weak energy (i.e. bending $E_b$) and the strong energy (i.e. stretching $E_s$) \cite{lobkovsky,conti2008confining}. This idea, motivated by related results in statistical and classical mechanics, goes by various names -- `equipartition', `dominant balance' and `virial theorem'. However, the occurrence of singularly perturbed Euler-Lagrange equations is by itself not a guarantee for equipartition between the bending and stretching energies or the occurrence of multiple scale behavior. An illustrative example is the comparison between a thin elastic rod that is confined by a ring in two dimensions, and a thin elastic sheet that is confined by a sphere\footnote{This example was suggested by Tom Witten}. In contrast with the crumpling that occurs in a sheet, the rod ``curls up'' parallel to itself touching the boundary with a uniform curvature on the same scale as the forcing, i.e. the curvature of the confining ring. Moreover, for the rod $E_s/E_b \to 0$ as the thickness vanishes, so this system does not display equipartition.  Nevertheless, for both the rod and the sheet, the ratio of the bending to the stretching stiffness is proportional to the square of the thickness $t$, but the resulting configurations are very different. This example shows that the presence of multiple energy scales, does not in itself create multiple scale behavior. The geometry of the system plays a significant role in determining the structures that arise spontaneously. 

The preceding example is often explained away as an exceptional case, namely that equipartition doesn't hold because the solutions themselves are not ``multi-scale". However, Davidovitch et al.  pursue the idea that rather than being an exception, this failure of equipartition, $E_s/E_b \to 0$  is actually a widespread feature, that they dub the Gauss-Euler elastica \cite{Davidovitch2019Geometrically}. They ``turn equipartition around" by demanding that $E_s/E_b \to 0$ and use the Gauss-Euler elastica principle as a quantitative tool to calculate multi-scale configurations of thin sheets. In this setting, the ``limiting" states are examples of {\em asymptotic isometries}. These ideas hold valuable lessons for physicists working on energy driven self-organization. Indeed, the idea that equipartition is not universal is not widely recognized. On the contrary, explanations for multiple scale phenomena based on equipartition, for example the results from~\cite{audoly2003self} on self-similar buckling in hyperbolic sheets~\cite{sharon2002buckling}, are considered robust and are thus deeply rooted in the physics community even in light of analyses of the phenomenon that lead to lower energy configurations and no equipartition~\cite{EPL_2016}.

\subsection{Non-Euclidean elasticity, energy scales and weak forces}

With the prior discussion serving as a backdrop, we discuss the roles that energy scales and weak forces play in determining the configuration of minimizers in the non-Euclidean modeling framework. We let $(x,y) \in \Omega$, a subset of $\mathbb{R}^2$, denote material coordinates on the center surface of free elastic sheet with a growth induced Riemannian metric $\mathbf{g}$. In this coordinate system the intrinsic distance between such points is given by the arc-length element:
\begin{equation*}
ds^2=g_{11}(x,y)dx^2+2g_{12}(x,y)dxdy+g_{22}(x,y)dy^2.
\end{equation*}
By the Kirchhoff hypothesis \cite{solid-mech-book}, the equilibrium configuration is modeled by an  immersion $F:\Omega\mapsto \mathbb{R}^3$ which minimizes an elastic energy  energy consisting of stretching and bending contributions:
\begin{align}
\label{elastic-energy}
E[F] &=\mathcal{S}[\gamma]+t^2\mathcal{B}[F]  \\
&=\int_{\Omega} Q(\gamma)\,dxdy+t^2 \int_{\Omega}(4H^2-2K)\,dxdy, \nonumber
\end{align}
where $\gamma = (\nabla F)^T\cdot \nabla F-\mathbf{g}$ denotes in-plane strains, $t$ is the thickness of the sheet, $Q(\gamma)=Q(\tr(\gamma),\det(\gamma))$ is a quadratic form, and $H$ and $K$  are respectively the mean and Gaussian curvatures of the center surface \cite{efrati2009elastic, lewicka2011foppl}. 

In this model, the ``thinness'' of an elastic sheet is reflected in the ratio of the flexural and in-plane rigidities of the sheet, i.e,  they are very easy to bend but much harder to stretch. Therefore, at least heuristically, we expect that in the vanishing thickness limit the sheet adopts a configuration minimizing bending energy while satisfying a zero in-plane stretching constraint, i.e. adopting an isometric immersion of $\mathbf{g}$. Stated more precisely, the $t \to 0$ (vanishing thickness) limit of minimizers of the elastic energy (\ref{elastic-energy}) minimize the bending ($W^{2,2}$) energy among all \emph{isometric immersions} $F:(\Omega,\mathbf{g}) \to \mathbb{R}^3$ \cite{lewicka2011scaling}. Rather than use the (technical) notion of $W^{2,2}$ isometric immersions, we define
{\em (piecewise smooth) isometries with finite bending content} as configurations $F$ satisfying:
\begin{enumerate}
    \item $\nabla F$ is continuous,
    \item $(\nabla F)^T\cdot \nabla F=\mathbf{g}$,
    \item ~$D^2F$, the Hessian of $F$, is piecewise smooth,
    \item $\int_{\Omega}(4H^2-2K)\,dx\,dy < \infty$.
\end{enumerate}
Such configurations are indeed $W^{2,2}$ isometries according to the precise mathematical definition. To illustrate the physical import of this definition, we note that the vanishing thickness limits of the microstructure that arises in crumpled sheets, i.e. elastic ridges \cite{lobkovsky} and $d$-cones \cite{benAmar1997Crumpled}, violate the first condition. They are neither in this class, nor in $W^{2,2}$; indeed the bending content of these defects diverges as $t \to 0$.

One benefit of studying this problem within the class of isometric immersions with finite bending content is that it simplifies the analysis when including weak external forces. As a specific example, the gravitational potential energy $$
\mathcal{G} = \rho g \int_\Omega F\cdot \mathbf{e}_3 dx dy,
$$
may be included into this model as a prototypical example of a \emph{weak external force} {\em provided that $\rho g$ scales as $t^2$}. The full problem requires minimizing all three energies, i.e., stretching, bending, and gravity, but is difficult to solve. For the particular physical scaling $\rho g \sim t^2$,  however, the problem is in an asymptotic regime in which the $t\to 0$ limit configurations are isometric immersion with finite bending content. Specifically, for free, thin sheets where the external forces, e.g., gravity, is weak and scales similarly as bending, the energy minimizers can be approximated by minimizers of the gravitational and bending energies over the class of exact isometries with finite bending content. 

There are also other interesting asymptotic regimes in which external forces are not weak and scale differently than bending, e.g., in stamping problems investigated in \cite{Aharoni2017Smectic,Albarrn2018Curvature,Tobasco2020Principles,Tobasco2021Curvature} for non-Euclidean shells floating on top of a planar liquid surface and in \cite{Hure2012Stamping,Davidovitch2019Geometrically,King2012Elastic,bella2014wrinkles} for flat sheets confined onto the surface of a sphere. In these scenarios, the external forces are strong in that the `substrate' energy scales as $t^\alpha$ with $\alpha < 2$, leading to asymptotic, rather than true isometries.  However, when it can be proven that the limiting solution for an asymptotic family of energy-minimization problems is an exact isometry, there are advantages to assuming exact isometries a priori, e.g., simplifications to the analysis, using the isometry constraint to derive scaling laws, etc. Indeed a key result of the work presented in this paper is that for hyperbolic free sheets there is a third scaling regime, beyond equipartition and the Gauss-Euler elastica. In this regime, we get exact, rather than asymptotic isometries as the limiting states, but we can nonetheless see ``multi-scale" patterns that arise from a competition between the energetic contributions of the two principal curvatures in an isometric immersion \cite{EPL_2016,Shearman2021Distributed}. Stretching energy is now negligible. Instead of local equipartition between stretching and bending energy, we expect  that the separate contributions from the two principal curvatures will balance in an averaged sense \cite{Shearman2021Distributed}.

\subsection{A roadmap}

We close this introduction with a guide to the reader, highlighting the key results of this work. In \S\ref{S:GaM} we consider the equilibria of thin elastic sheets with intrinsic hyperbolic geometry in the small-slope setting. We show that exact isometries correspond to solutions of a fully nonlinear, hyperbolic PDE -- the Monge-Ampere equation. We show by explicit construction that $C^{1,1}$ isometric immersions are ``flexible" and can be assembled by piecing together ``elementary quads" that give local solutions of the Monge-Ampere equation. We are naturally led to the consideration of a novel type of defect, {\em branch points}, that arise in $C^{1,1}$ solutions of hyperbolic Monge-Ampere equations, and are ``robust" because they are associated with a topological invariant, the winding number of the normal to the surface. These defects are remarkably different from other defects in condensed matter systems in that they do not concentrate energy. Yet, they play a substantial role in the mechanics of thin hyperbolic sheets by circumventing the rigidity of smooth isometric immersions. They allow for a combinatorially large number of low energy states, thus making hyperbolic thin sheets a very ``floppy" system with fascinating mechanical properties.

In \S\ref{sec:applications} we present our results on how branch point defects contribute to the {\em extreme mechanics} of hyperbolic sheets. We highlight the role of branch points in (i) mediating morphological changes in the process of growth of thin hyperbolic laminae, (ii) actuating shape control, (iii) non-monotonic force-displacement relationships for confined hyperbolic sheets, (iv) responsiveness of hyperbolic sheets to weak external forces, and (v) potential applications to soft robotics for realizing large shape changes with small energy budgets. We conclude in \S\ref{sec:discussion} with a discussion of our  results. 

%%%%%%%%%%%%%%%%%%%%%%%%%%%%%%%%%%%%%%%%%%%%%%%%
%
%		Hyperbolic Geometry
%
%%%%%%%%%%%%%%%%%%%%%%%%%%%%%%%%%%%%%%%%%%%%%%%%

 \section{Geometry and mechanics}\label{S:GaM}

We begin with a discussion of the small-slopes or linearized geometry which describes elastic sheets with {\em small prestrain}, i.e. sheets whose metric is close to Euclidean (intrinsic curvature is small) and whose embedding into 3-space has small slopes relative to a ``nominal" horizontal plane $w=0$. The smallness of the slopes/deviation from the Euclidean metric is quantified by a parameter $\epsilon \ll 1$. 

The intrinsic geometry of the sheet is given by a metric, represented by the matrix
\begin{equation}
\mathbf{g} = \begin{bmatrix} 1+ \epsilon^2 f_{11} (x,y) & \epsilon^2 f_{12}(x,y) \\
\epsilon^2 f_{12}(x,y)   & 1+ \epsilon^2 f_{22}(x,y) \end{bmatrix},
\label{eq:g_metric}
\end{equation}
where $x,y$ are intrinsic coordinates on the sheet. The small-slopes embedding is given by the F\"oppl - von K\'arm\'an ansatz 
\begin{equation}
F(x,y) = \begin{bmatrix} x+\epsilon^2 \xi(x,y) \\ y+\epsilon^2\eta(x,y)  \\ \epsilon w(x,y) \end{bmatrix}, 
\label{FvK-embedding}
\end{equation}
with $O(\epsilon^2)$ in-plane and $O(\epsilon)$ out-of-plane deformations. The (extrinsic) metric induced by the embedding is $\mathbf{G} = dF \cdot dF$, given to $O(\epsilon^2)$ by 
$$
\mathbf{G} = \mathbf{I} +  \epsilon^2\begin{bmatrix} w_x w_x+ 2 \xi_x &  w_x w_y + \xi_y + \eta _x \\
w_x w_y + \xi_y + \eta _x & w_y w_y + 2 \eta_y \end{bmatrix},
$$ 
where $\mathbf{I}$ is the identity matrix. Small-slope isometries are given by the zero-strain condition $\mathbf{G} - \mathbf{g} = \mathbf{0}$, which corresponds to the system
\begin{align}
w_x w_x+ 2 \xi_x & = f_{11}(x,y), \nonumber \\
w_x w_y + \xi_y + \eta _x & = f_{12}(x,y), \nonumber \\
w_y w_y + 2 \eta_y & = f_{22}(x,y).
\label{eq:no_strain}
\end{align}
We can eliminate the in-plane displacement functions $u$ and $v$ by taking the curl along the rows and the columns of the strain matrix $\mathbf{G}-\mathbf{g}$ to obtain the Monge-Ampere equation
\begin{align}
w_{xx} w_{yy} & - (w_{xy})^2 = \nonumber \\
&  -\frac{1}{2} \left[\partial_{yy} f_{11} -2 \partial_{xy} f_{12} + \partial_{xx} f_{22} \right].
\label{eq:mng_amp1}
\end{align}

Eq.~\eqref{eq:mng_amp1} is a necessary and sufficient condition for the existence of a small-slopes isometry for the metric $\mathbf{g}$ in Eq.~\eqref{eq:g_metric}. If the right hand side, i.e. the intrinsic (linearized) Gaussian curvature 
$$ 
-\frac{1}{2} \left[\partial_{yy} f_{11} -2 \partial_{xy} f_{12} + \partial_{xx} f_{22} \right] := K_{\text{int}}(x,y)
$$ 
is positive everywhere, then Eq.~\eqref{eq:mng_amp1} is an elliptic Monge-Ampere equation \cite{trudinger2008monge}. Motivated in part by applications to optimal transport \cite{DePhilippis2014Monge}, there is a well developed existence, uniqueness and regularity theory for solutions of elliptic Monge Ampere equations ({\em cf.} \cite{Caffarelli1985Dirichlet}), along with numerical methods for these equations with convergence guarantees \cite{Benamou2010Two}. 

For the examples in Fig.~\ref{fig:examplesofnaturalobjects}, and more generally, for the systems of interest in this work, the intrinsic Gaussian curvature is negative, corresponding to hyperbolic elastic sheets. There exist ``special" solutions for these equations, for instance the product solution considered in \cite{EPL_2016}, but, there are impediments for a general mathematical theory for ``all" solutions of hyperbolic Monge-Ampere equations because one neither expects uniqueness, nor regularity for the Cauchy problem with smooth initial/boundary data.

\subsection{Small-slopes and constant curvature} \label{sec:small-slope}
We begin by first considering ``local" solutions near a point $x_0$, so that without loss of generality the Gauss-curvature $K_{\text{int}}(x,y)$ can be replaced by a constant $K_0 = K_{\text{int}}(x_0,y_0) = -1$ by picking the units of length appropriately. More generally, for the hyperbolic metric 
\begin{equation}
    \label{eq:constant-K-metric}
    \mathbf{g} = dx^2 + dy^2 + \frac{\epsilon^2}{3} (ydx-xdy)^2,
\end{equation}
a calculation using Eq.~\eqref{eq:mng_amp1} gives $K_{\text{int}} = -1$ so that the out-of-plane deformation for isometries is given by $\epsilon w(x,y)$ where 
\begin{equation}
    \label{eq:mng_amp_cnst}
    \mathrm{det}(D^2 w) = w_{xx} w_{yy} - w_{xy}^2 = -1.
\end{equation}
Assuming that $w(x,y)$ is a smooth solution to Eq.~\eqref{eq:mng_amp_cnst}, we get the Taylor expansion 
\begin{align*}
w(x,y)  = & w_0 + p_0(x-x_0) + q_0 (y-y_0) \nonumber \\
& + \frac{1}{2} \begin{bmatrix} x-x_0 & y-y_0\end{bmatrix} Q  \begin{bmatrix} x-x_0 \\ y-y_0\end{bmatrix} + \text{h.o.t.},
\end{align*}
where  $Q = D^2 w(x_0,y_0)$ is a symmetric $2 \times 2$ matrix with $\mathrm{det}\,Q = -1$ and arbitrary $w_0$,$p_0$, and $q_0$. In particular, $Q$ can be diagonalized by a special orthogonal transformation $O$, i.e. $O^T O = \mathbf{I}, \mathrm{det}(O) = 1$, so that for some $a > 0$,  
\begin{align}
\label{in_asymp_coords}
Q & = O^T \begin{bmatrix} a^2 & 0 \\ 0 & -a^{-2} \end{bmatrix} O   \\
& = (A O)^T \begin{bmatrix} 0 & 1 \\ 1 & 0 \end{bmatrix} (A O), \nonumber \end{align}
where
\begin{align}
    A & = \frac{1}{\sqrt{2}}\begin{bmatrix} a & a^{-1} \\ a & -a^{-1} \end{bmatrix}, \nonumber
\end{align}
and thus $\det(A)=-1$. 
\begin{figure}[htbp]
\begin{center}
\includegraphics[width=0.95\linewidth]{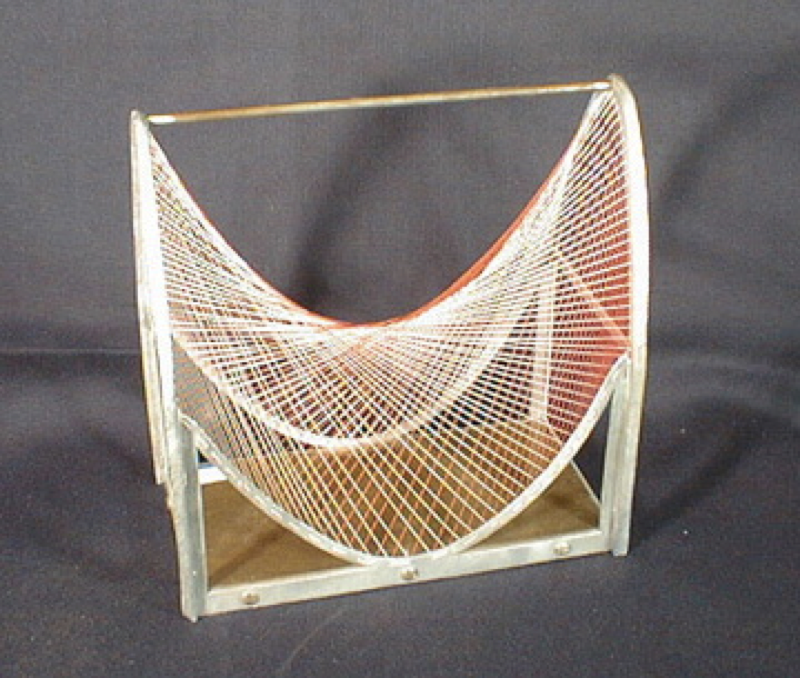}
\caption{A quadratic saddle surface is ruled by two families of straight lines. This image is of a model displayed at the University of Arizona Mathematics department.}
\label{fig:ruled}
\end{center}
\end{figure}

We define the linear forms $l_1$ and $l_2$ by 
$$
\begin{bmatrix} l_1(x,y) \\ l_2(x,y) \end{bmatrix} = A O \begin{bmatrix} x-x_0 \\ y-y_0\end{bmatrix}
$$
which gives 
\begin{align*}
w(x,y)   = & w_0 + p_0(x-x_0) + q_0 (y-y_0) \\
& + l_1(x,y) l_2(x,y).
\end{align*}
In particular, along the lines $l_1 = $ constant or $l_2 = $ constant, $w$, $x$ and $y$ are linear functions of an affine parameter along the line. Consequently, a quadratic surface with negative curvature, i.e. a quadratic saddle, is ruled by two families of straight lines as illustrated in Fig.~\ref{fig:ruled}. This surface decomposes into a collection of {\em quads} (i.e. skew parallelograms/rhombi) given by the two families of rulings, and the ``local" solution we seek is therefore the restriction of this quadratic surface to a single quad. A natural idea is to build ``global" solutions for the Monge-Ampere equation by assembling such quad elements.

\begin{figure}[htbp]
\begin{center}
 \begin{subfigure}[b]{\linewidth}
 \centering
\includegraphics[trim={1cm, 1cm, 1cm, 1.3cm}, clip,width=0.8\linewidth]{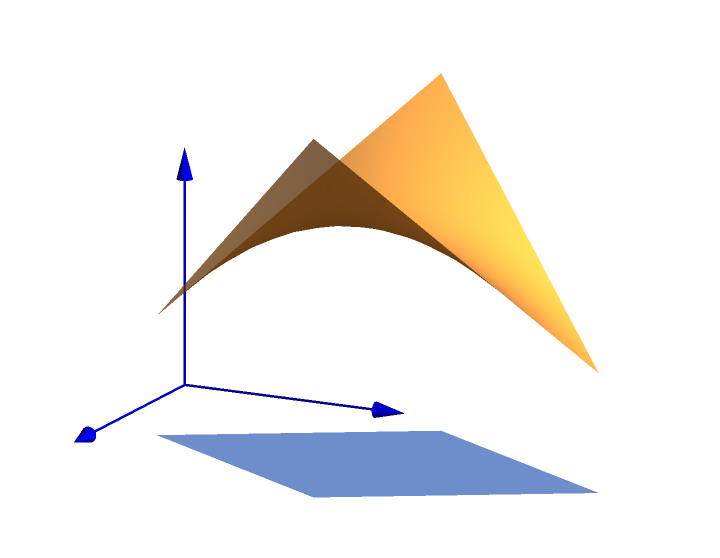}
  \caption{$xy$ projection.}
 \label{fig:project}
\end{subfigure}
 \begin{subfigure}[b]{\linewidth}
 \centering
\includegraphics[width=0.4\linewidth]{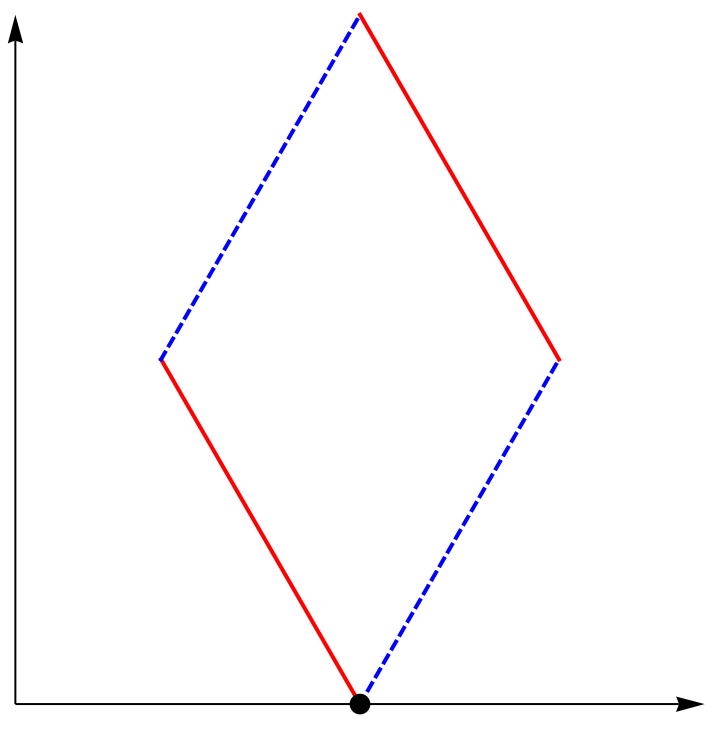} \quad
\includegraphics[width=0.4\linewidth]{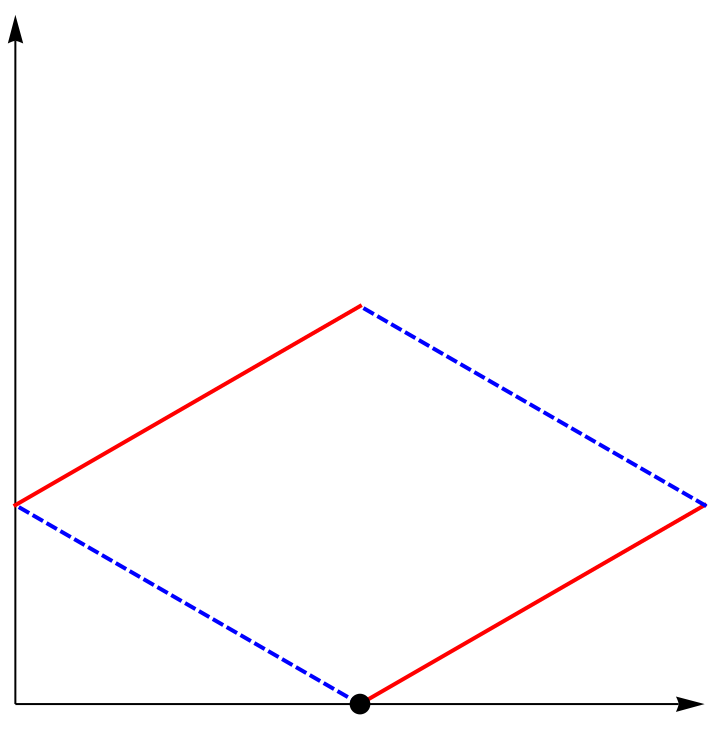}
  \caption{Dual rhombi.}
 \label{fig:dual}
\end{subfigure}
\caption{(a) The $xy$-projection of an elementary quad in a saddle, bounded by edges from the rulings, gives a rhombus. (b) The $xy$ and $pq$-projections of an elementary quad. The $u$-edges are the solid lines and the $v$-edges are the dashed lines, and the nodes corresponding to a given point $r_0$ are indicated by the small disks. The dual rhombus, given by projection on to the gradient $(p,q) = \nabla w$, has the opposite orientation and edges which are orthogonal to the corresponding edges in the $xy$ projection.}
\label{fig:quad}
\end{center}
\end{figure}

In the neighborhood of the point $(x_0,y_0)$ we define the {\em asymptotic coordinates} by $u = l_1(x,y)$ and $v = l_2(x,y)$ so that we have the parametric representation
\begin{align}
\begin{bmatrix} x \\ y \end{bmatrix} & = \begin{bmatrix} x_0 \\ y_0 \end{bmatrix} + (AO)^{-1} \begin{bmatrix} u \\ v \end{bmatrix}, \nonumber \\
w & = w_0 +  \begin{bmatrix} p_0 & q_0 \end{bmatrix} (AO)^{-1} \begin{bmatrix} u \\ v \end{bmatrix} + u v, \nonumber \\
\nabla w & := \begin{bmatrix} p \\ q \end{bmatrix} =   \begin{bmatrix} p_0 \\ q_0 \end{bmatrix} + (AO)^{T} \begin{bmatrix} v \\ u \end{bmatrix}. 
\label{eq:parametric}
\end{align}
It follows from these formulae that on every ``elementary quad", which we can associate with the square grid cell $0 \leq u \leq \ell$, $0 \leq v \leq \ell$ for some small side-length $\ell$, the Lagrangian coordinates $(x,y)$ and the ``hybrid" gradient of the Eulerian out-of-plane deformation $\nabla w$ with respect to the Lagrangian coordinates $(x,y)$, are both linear functions of $(u,v)$. Indeed, more is true. By considering differences of these quantities at the point $(u_1,v_1)$ and $(u_2,v_2)$, we see that   
\begin{equation}
\begin{bmatrix} \Delta p  &  \Delta q \end{bmatrix} \begin{bmatrix} \Delta x \\ \Delta y \end{bmatrix} = 2 \Delta u \cdot \Delta v,
\label{consistency}
\end{equation}
where $\Delta f := f_2-f_1$ denotes the difference operator acting on the quantity $f$. In particular, if $\Delta u$ or $\Delta v$ is zero, then it follows that the vector $\Delta(p,q)$ is orthogonal to the corresponding vector $\Delta(x,y)$.
More specifically, using Eqs.~\eqref{in_asymp_coords}, \eqref{eq:parametric} and $O^T = O^{-1}$, we have
\begin{align}
\begin{bmatrix} \Delta p  \\  \Delta q \end{bmatrix} & =  \begin{bmatrix} \Delta y \\ -\Delta x \end{bmatrix} \mbox{ if } \Delta v=0, \nonumber\\ 
\begin{bmatrix} \Delta p  \\  \Delta q \end{bmatrix} & =  \begin{bmatrix} -\Delta y \\ \Delta x \end{bmatrix} \mbox{ if } \Delta u=0. 
\label{eq:dualize}
\end{align}
These relations are illustrated in Fig.~\ref{fig:quad}(b). We will henceforth refer to the edges with $\Delta v =0$ as the $u$-edges, and the edges with $\Delta u = 0$ as the $v$-edges and follow the standard convention in discrete differential geometry and use the subscripts $0,1,2$ and $12$ to denote the points corresponding to the $(u,v)$ coordinates $(0,0),(\ell,0),(0,\ell)$ and $(\ell,\ell)$ respectively. Letting $r = \begin{bmatrix} x\\y \end{bmatrix}$ and $\zeta =  \begin{bmatrix} p\\q \end{bmatrix}$, we have the (oriented) angle relation
\begin{equation}
\angle(\zeta_1 \zeta_0 \zeta_2) = \angle(r_1 r_0 r_2) - \pi
\label{eq:dual-angles}
\end{equation}
and the oriented areas of the quads in $r$ and $\zeta$ are related by 
\begin{equation}
\mathrm{area}(\zeta_0 \zeta_1 \zeta_{12} \zeta_2) =- \mathrm{area}(r_0 r_1 r_{12} r_2) 
\label{eq:weak-formulation}
\end{equation}
This last relation is the integrated (i.e. ``weak") form of the Monge-Ampere equation $\mathrm{det}(D^2w) = -1$ on a single quad bounded by asymptotic curves \cite{stoker}. The rulings are continuous families of lines, and we can, in general, pick representatives such that their projections are equi-spaced in the $xy$-plane so that the projections of the elementary quads are rhombi, as illustrated in Fig.~\ref{fig:quad}(a). We will henceforth make this choice.

%%%%%%%%%%%%%%%%%%%%%%%%%%%%%%%%%%%%%%%%%%%%%%%%
%
%		Asymptotic skeleton
%
%%%%%%%%%%%%%%%%%%%%%%%%%%%%%%%%%%%%%%%%%%%%%%%%

\subsection{The asymptotic skeleton} \label{sec:foliations}

\begin{figure}[htbp]
\begin{center}
\includegraphics[trim={1cm, 1.5cm, 1cm, 2cm}, clip,width=0.8\linewidth]{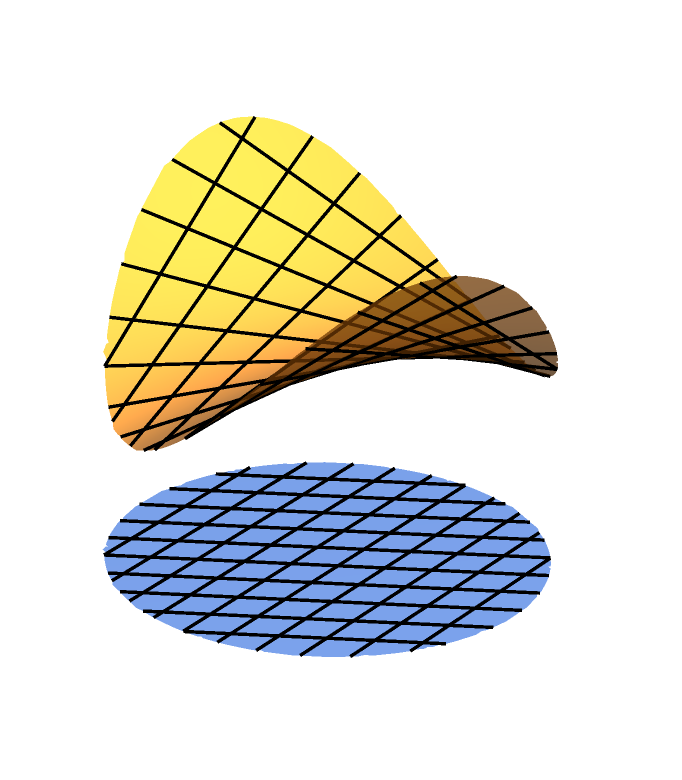}
\caption{The projection of the asymptotic lines that rule a quadratic saddle surface give a quadmesh consisting of congruent rhombi.}
\label{fig:quad-proj}
\end{center}
\end{figure}

We will now construct solutions of the Monge-Ampere equation $\mathrm{det}(D^2w) = -1$ by patching individual quad elements described by Eq.~\eqref{eq:parametric}. As we discuss above, the projection of the skew-quads in the ruled quadratic surface in Fig.~\ref{fig:quad}  to the $xy$-plane gives a family of parallelograms, which, with an appropriate spacing between the rulings, can be chosen to be rhombi. Further, as illustrated in Fig.~\ref{fig:quad-proj}, four rhombi meet at every interior node.

\begin{figure}[hbpt]
        \begin{subfigure}[b]{0.65\linewidth}
                \centering
                \includegraphics[width=\linewidth]{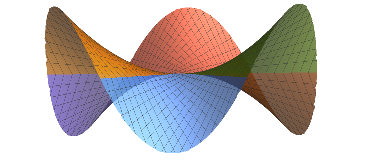}
                \caption{Piecewise quadratic}
                \label{fig:mnky-saddle}
        \end{subfigure}%
         \begin{subfigure}[b]{0.35\linewidth}
                \centering
                \includegraphics[width=\linewidth]{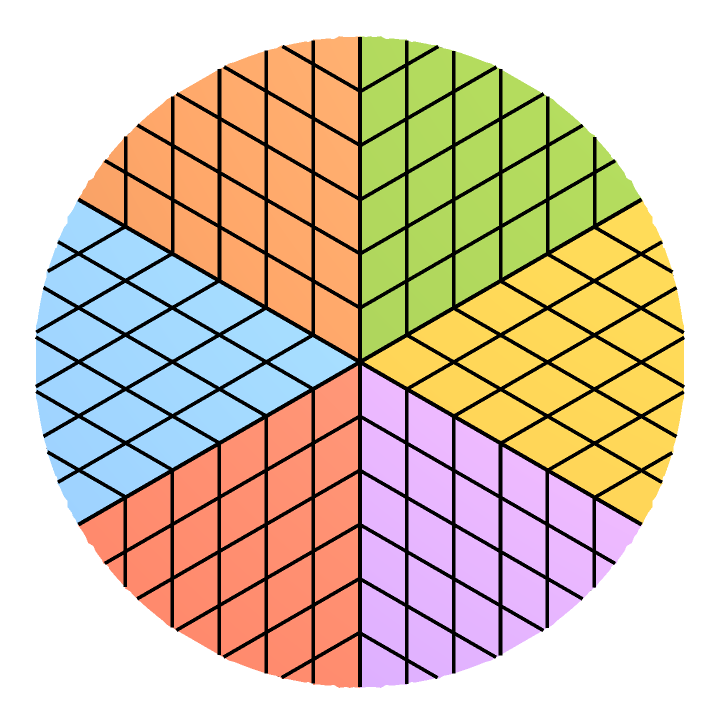}
                \caption{Projection}
                \label{fig:a-skeleton}
        \end{subfigure}%
 
        \caption{}
        \label{fig:patching}
\end{figure}

We can build piecewise quadratic surfaces by relaxing the condition that only four rhombi meet at every interior node. We illustrate with an example -- the saddle surface $w = \frac{x^2}{\sqrt{3}} - \sqrt{3} y^2$ projects down to the rhombi with angles $\pi/3$ and $2 \pi/3$. Consequently, we can build a piecewise quadratic surface by patching together 6 copies of the sector $x \geq 0, |y| \leq \frac{x}{\sqrt{3}}, w = \frac{x^2}{\sqrt{3}} - \sqrt{3} y^2$ by odd reflections, as illustrated in Fig.~\ref{fig:patching}. In this case, the origin is incident on six rhombi, while all the other interior nodes are incident on four rhombi.

To each rhombus in the $xy$-projection, we can associate a ``dual" rhombus in the $pq$-projection, i.e. in ``gradient space", through Eq.~\eqref{eq:dualize}. Since the $u,v$-edges alternate on each rhombus, and the $u,v$-labels have to agree on the edges common to adjacent faces, it follows that an even number $2m_i \geq 4$ are incident on every interior node $r_i$. Eq.~\eqref{eq:dual-angles} now implies that the sum of the angles of the dual rhombi incident on an interior vertex $\zeta_i$ is related to the sum of the angles of the rhombi incident on $r_i$ by 
$$
\sum \text{dual angles} = 2 (1-m_i) \pi,
$$
where we use the fact that, in the $xy$-projection, the angles of the rhombi incident on an interior node should add up to $2 \pi$. 

\begin{figure}[ht]
\centering
        \begin{subfigure}[t]{0.48\linewidth}
                \centering
                \includegraphics[trim={4.75cm 5.75cm 4cm 4.5cm}, clip, width=\linewidth]{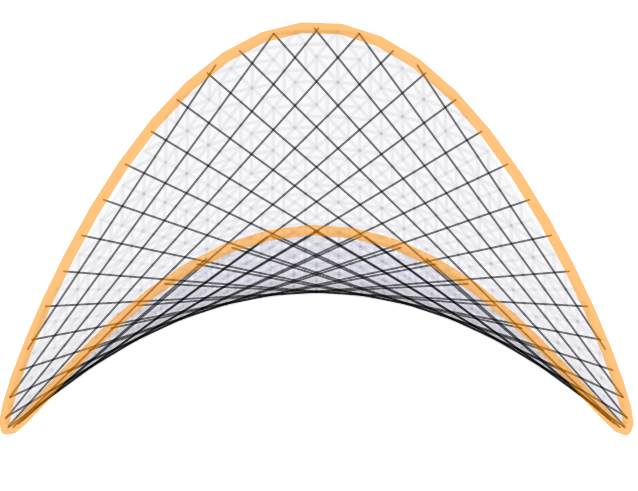}
                \caption{}
                \label{fig:saddle}
        \end{subfigure}%
        \begin{subfigure}[t]{0.48\linewidth}
                \centering
                \includegraphics[trim={0cm -1.25cm 0cm 0cm}, clip, width=\linewidth]{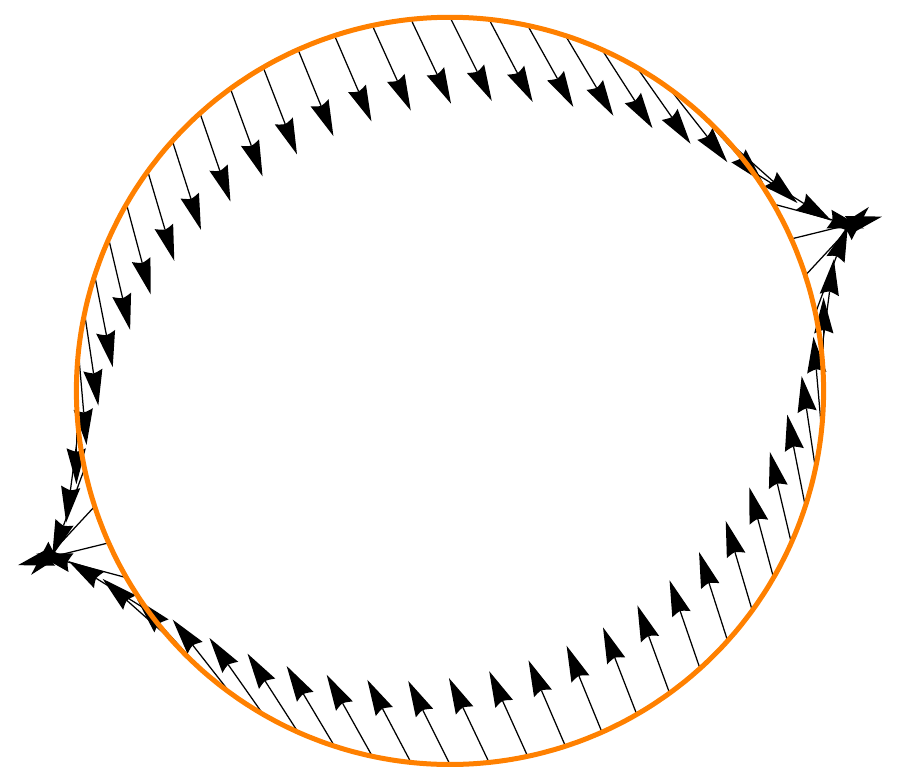}
                \caption{}
                \label{fig:nrmlsaddle}
        \end{subfigure}
        
        \begin{subfigure}[t]{0.48\linewidth}
                \centering
                \includegraphics[trim={5cm 4.5cm 4.25cm 6.cm}, clip, width=\linewidth]{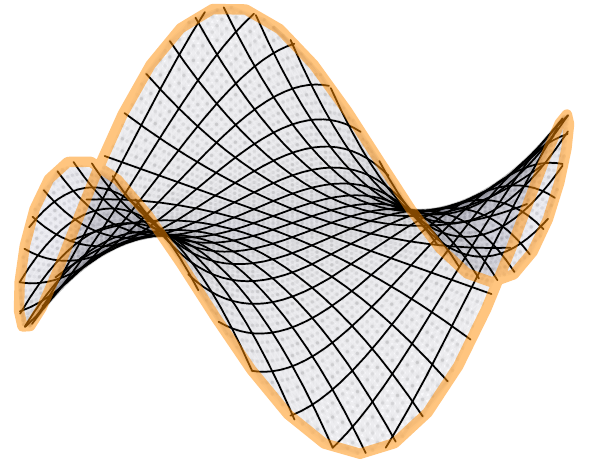}
                \caption{}
                \label{fig:mkysaddle}
        \end{subfigure}
        \begin{subfigure}[t]{0.48\linewidth}
                \centering
                \includegraphics[width=\linewidth]{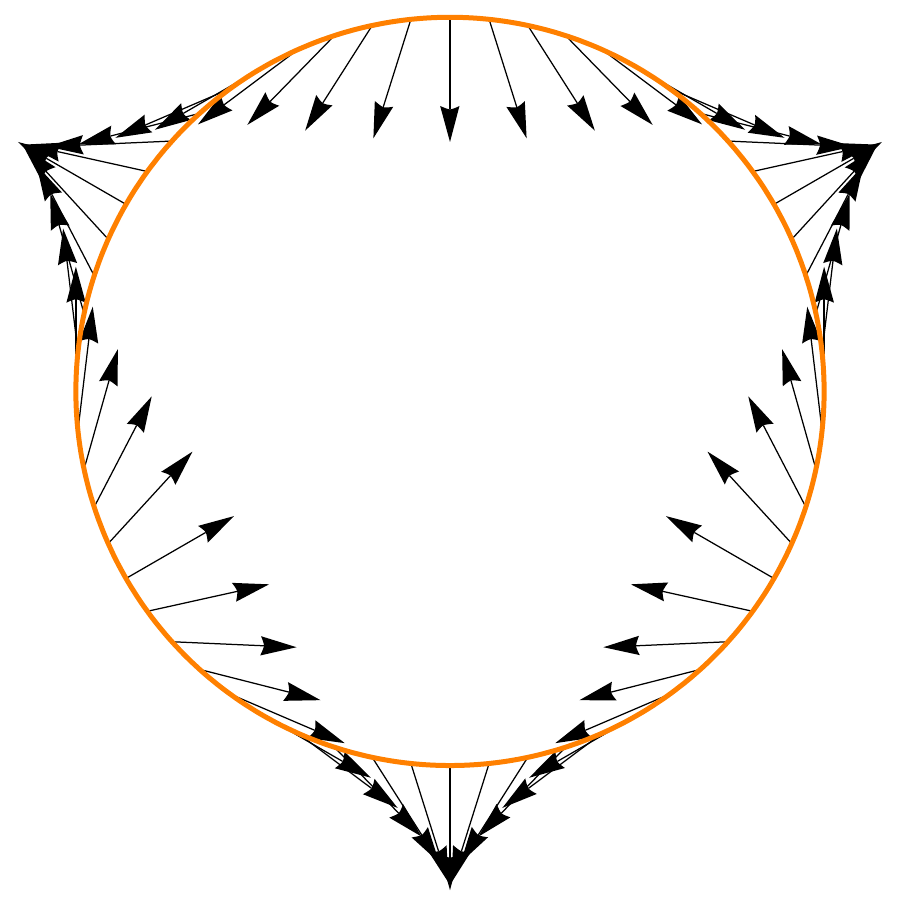}
                \caption{}
                \label{fig:nrmlmkysaddle}
        \end{subfigure}        
        \caption{The (local) winding number of the gradient field about a point $p$ for two surfaces: (a) A smooth saddle and (c) A monkey saddle with a branch point at the origin. The mesh is obtained by lifting a regular square mesh on the $xy$ plane to the surface, and does not correspond to the asymptotic quads, in contrast to Fig.~\protect{\ref{fig:patching}}. (b) and (d) are projections of the corresponding gradients $\nabla w$ along the boundary, which is a circle enclosing the branch point in (a) and (c), respectively. With respect to the centers of the disks, $m = 2$ for the saddle and $m = 3$ for the monkey saddle.}\label{fig:winding}
\end{figure}

We will refer to a node $r_i$ incident on four rhombi, i.e. the nodes with $m_i=2$, as a {\em regular node}. Traversing a small circle $r(\theta) = r_0 + \delta e^{i \theta}$, centered on the node $r_0$ in a counter-clockwise direction, the gradient $\zeta(\theta)$ will traverse a simple closed curve in the clockwise direction, since the total angle is $-2\pi$. For nodes with $m_i > 2$, however, the gradient $\zeta$ will traverse a {\em non-simple} closed curve that winds around $\zeta(r_0)$ a total of $m_i-1$ times, again in a clockwise sense. The nodes with $m_i > 2$ are therefore {\em branch points} for the gradient map $(x,y) \mapsto \nabla w(x,y)$, and in particular, the gradient map is not differentiable at these branch points since the image of the map is multi-sheeted and cannot be approximated locally by a linear map. This is illustrated in Fig.~\ref{fig:winding}.  

From the above considerations, we arrive at the following algorithm for constructing small-slope, no-stretching, piecewise smooth solutions of Eq.~\eqref{eq:mng_amp_cnst}, i.e. $\mathrm{det}(D^2w) = -1$ with finite bending content, i.e. in the class $W^{2,2}$.
\begin{enumerate}
    \item Given a domain $\Omega \subset \mathbb{R}^2$, and a discretization (length) parameter $\ell$, by considering a larger set $D \supset \Omega$ if necessary, find a {\em quadgraph} \cite{Bobenko2008Discrete,Huhnen-Venedey2014Discretization} covering $D \supset \Omega$, i.e. a cellular decomposition \cite{HatcherAlgTop} of $D$ into rhombi with side $\ell$ such that all the interior vertices have even degrees. 
    \item Construct the dual rhombi using Eq.~\eqref{eq:dualize}. This gives, in general, a branched covering, i.e. a ``multi-sheeted" subset of the $pq$ gradient plane decomposed into rhombi.
    \item Use the formulae in Eq.~\eqref{eq:parametric} to construct a piecewise surface. By construction, the gradients on all the pieces incident on a common vertex agree at this vertex. Along the edges that are common to adjacent rhombi, the gradients are given by linear functions that agree at the endpoints of the edge, so they have to agree everywhere along the edge. Consequently we get a piecewise smooth (on each rhombic face) and globally differentiable surface. 
\end{enumerate}

\begin{figure}%[tbhp]
\centering
\includegraphics[width=\linewidth]{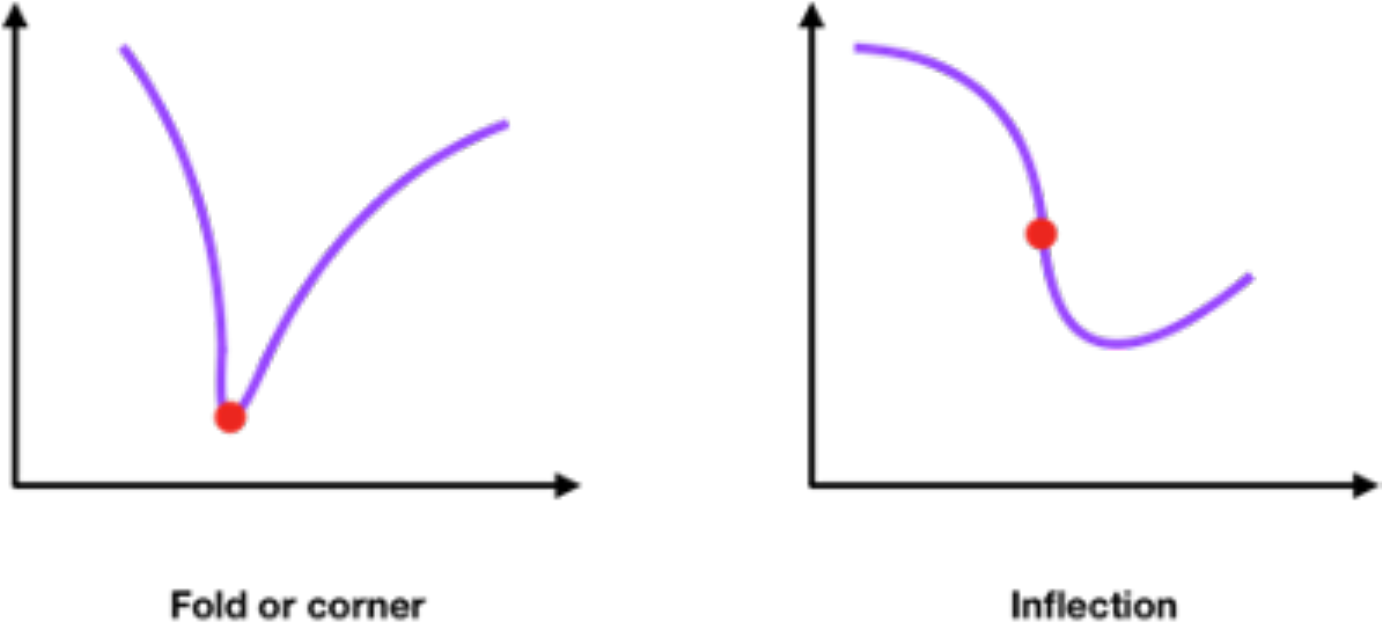}
\caption{The contrast between an ``edge" in a crumpled flat sheet and a piecewise smooth hyperbolic sheet. In a crumpled sheet, the edges correspond to jumps in the gradient $\nabla w$, so the configuration is not $W^{2,2}$. On the other hand, the edges in the asymptotic skeleton of a $C^{1,1}$ hyperbolic isometry are lines of inflection and correspond to jumps in the second derivative $D^2 w$, while the gradient $\nabla w$ remains continuous. Such edges {\em do not} concentrate bending energy.}
\label{fig:C0vsC11}
\end{figure}

We will call the collection of all the edges of the elementary quads the {\em asymptotic skeleton} of the surface, since the edges give a discretization of the asymptotic curves, i.e the curves with zero normal curvature \cite[Chap. IV]{stoker}. Consequently, the edges between adjacent faces are {\em lines of inflection} and the resulting surfaces are $C^{1,1}$, but in general not $C^2$, as illustrated in Fig.~\ref{fig:C0vsC11}. Indeed, for the surface to be $C^2$ in the neighborhood of a vertex, we need that the surface be locally quadratic (and not just piecewise quadratic) implying that the vertex should have degree  four and, as in Fig.~\ref{fig:quad-proj}, the two $u$-edges (resp. $v$-edges) incident on the vertex should not form a kink.

\begin{figure}[htbp]
\begin{center}
\includegraphics[width=0.95\linewidth]{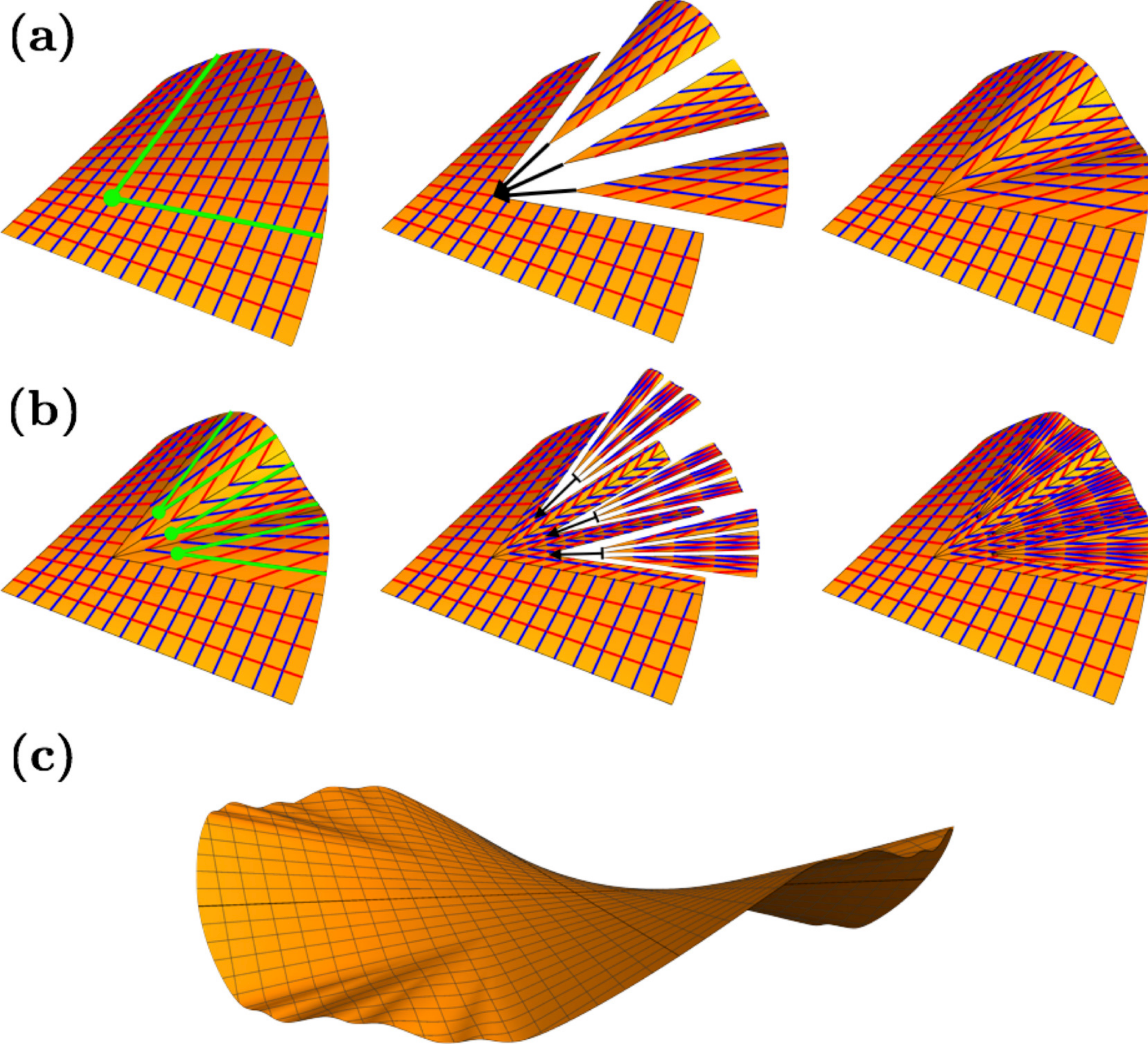}
\caption{Solutions to $\det(D^2 w)=-1$ with branch points. (a) Three sub-wrinkle solution on the first quadrant. (b) Nine sub-wrinkle solution.  (c) Extension to 36 sub-wrinkle solution.}
\label{fig:SelfSimilar}
\end{center}
\end{figure}

This construction shows that there are large families of  weak $W^{2,2}$ solutions of $\mathrm{det}(D^2 w) = -1$ given by tilings of (subsets of) $\mathbb{R}^2$ by rhombi with the additional condition that the degree of the resulting edge graph is even at each interior vertex. These conditions allow for the possibility of ``surgery" as shown in Fig.~\ref{fig:SelfSimilar}(a), where a sector is excised and replaced by three sectors built from rhombi of the same side length. This process introduces a branch point as well as lines of inflection into the surface. The surfaces in the first and the third panel in Fig.~\ref{fig:SelfSimilar}(a) agree on a neighborhood of the ``boundary" but are not equal, showing the lack of uniqueness and of regularity for weak solutions of $\mathrm{det}(D^2 w) = -1$. 

We can deform the location of the branch point, as well as the angles of the ``additional" sectors at the branch points in a continuous manner, showing that we have ``extreme" nonuniqueness of solutions and multi-parameter continuous families of weak-solutions of $\mathrm{det}(D^2 w) = -1$ that correspond to zero-stretching, finite bending content small-slope isometries with prescribed boundary conditions. As a consequence, hyperbolic elastic sheets are extremely floppy due to continuous families of low energy states.

Recursive surgery introduces multiple branch points on the surface,
as illustrated in Fig.~\ref{fig:SelfSimilar}(b). 
The resulting surface is not smooth, 
but it has a continuous tangent plane, bounded principal curvatures, and finite bending energy density. It also displays {\em refinement} or {\em sub-wrinkling} -- the ``number of waves" increases with $r$. 

While we have discussed these properties in the context of small-slope solutions with constant curvature, they have direct analogs for pseudospherical surfaces, i.e. surfaces whose ``full" curvature is equal to a negative constant. Such surfaces can be built from elementary rhombi \cite{sauer1950parallelogrammgitter,Wunderlich1951Differenzengeometrie} whose adjacency relations are encoded in an asymptotic skeleton \cite{Shearman2021Distributed}. These surfaces also support branch points that confer a great degree of flexibility to the class of $W^{2,2}$ solutions and result in pseudospherical elastic surfaces being mechanically floppy \cite{Shearman2021Distributed}.

The analog of~\eqref{eq:dualize} for the full (not necessarily small-slopes) geometry, on an elementary quad, is the system \cite{sauer1950parallelogrammgitter,Wunderlich1951Differenzengeometrie}
\begin{align}
\Delta \mathbf{r} & =  -\Delta \mathbf{N} \times \mathbf{N} \mbox{ if } \Delta v=0, \nonumber\\ 
\Delta \mathbf{r} & =  \Delta \mathbf{N} \times \mathbf{N} \mbox{ if }\Delta u=0. 
\label{eq:ddg_lelieuvre}
\end{align}
where $\mathbf{r}_i \in \mathbb{R}^3$ is a discretization of the surface and $\mathbf{N}_i \in \mathbb{S}^2$ is the corresponding unit normal. These equations reduce to the small-slopes equations~\eqref{eq:dualize} by ``linearizing" $$
\mathbf{N} \approx \hat{\mathbf{e}}_z - p  \hat{\mathbf{e}}_x  -  q \hat{\mathbf{e}}_y , \quad |\Delta \mathbf{r}| \sim |\Delta \mathbf{N}| \sim O(\delta).
$$  

%%%%%%%%%%%%%%%%%%%%%%%%%%%%%%%%%%%%%%%%%%%%%%%%
%
%		Applications
%
%%%%%%%%%%%%%%%%%%%%%%%%%%%%%%%%%%%%%%%%%%%%%%%%

\section{Mechanics, growth, and dynamics of hyperbolic non-Euclidean plates} \label{sec:applications}

In this section we discuss the mechanical and conformational responses of thin hyperbolic sheets to external or internal stresses, with an emphasis of the role that the branch points play in mediating various processes including shape changes, nonlinear mechanical responses and growth. 

We begin in \S\ref{sec:leaves} with an unconstrained geometric problem, the construction of isometries of negatively curved disks with increasing radii, that models the growth of a hyperbolic sheet, for example a leaf. We show that the growth process naturally leads beyond small-slope theory, and the nonlinear behavior is modulated by the appearance and the movement of branch points in the sheet. We then consider constrained geometric problems, and investigate the actuation of shape changes by controlling the locations of branch points (\S\ref{sec:shapecontrol}), as well as the response of the sheet to geometric confinement (\S\ref{sec:expts}). In \S\ref{S:WeakForcesScaling}
 we highlight the floppiness of thin hyperbolic sheets by obtaining scaling laws describing the energy and the geometry of these sheets in an external weak potential. The final application that we discuss, in \S\ref{sec:rotate}, is for soft robots. Here, we develop a novel framework for formulating the mechanics of a thin hyperbolic sheets with (potential) branch points interacting with an environment. An additional twist is that, in contrast to the earlier examples, we also need to correctly incorporate in-plane displacements in the analysis.

\subsection{Growing leaves and distributed branch points}\label{sec:leaves}

A natural extension of the discussion in the previous section is to consider problems where the geometry and/or the domain $\Omega(t)$ is changing with time, a process that one might call {\em growth}. As in the case of static sheets, finite-bending isometries are energetically preferable to non-isometric configurations in growing thin sheets at each moment in time. An idealized example of a growing sheet is  a ``circular leaf" given by 
a hyperbolic disk with radius $R(t) = t$. In the small-slopes approximation, the quadratic saddle $z = \frac{1}{2}(x^2 - y^2), x^2+y^2 \leq t^2$ is a smooth `isometric' solution of Eq.~\eqref{eq:mng_amp_cnst} for all times $t$.

However, the maximum slope for this solution grows with $t$ so, for sufficiently late times, it is no longer justified to use the small-slope approximation. Using the discrete equations~\eqref{eq:ddg_lelieuvre} for the full geometry, we can track the shape of a growing sheet beyond the small-slope regime. As the sheet grows, the introduction of branch points will reduce the bending energy and refine its wrinkling pattern. 

\begin{figure}[bpth]
\center
        \begin{subfigure}[b]{0.4\linewidth}
                \includegraphics[trim={1.cm, 0cm, 0.65cm, 6cm}, clip, width=\linewidth]{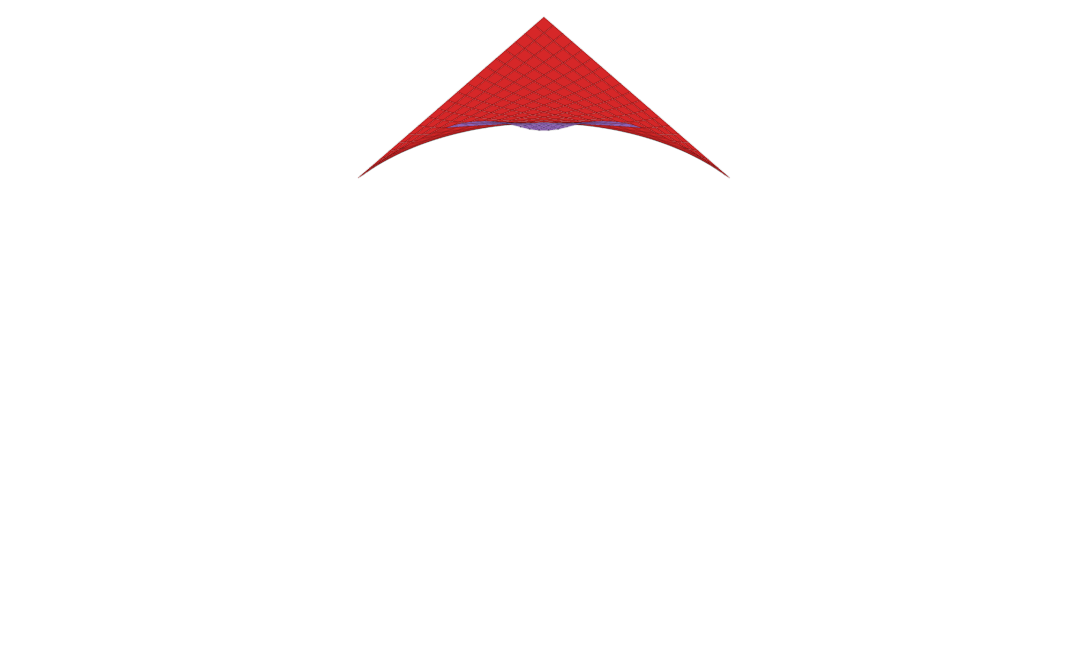}
                \caption{$r=1.15$}
        \end{subfigure}%
        \begin{subfigure}[b]{0.55\linewidth}
                \includegraphics[trim={1.cm, 0cm, 0.65cm, 6cm}, clip, width=\linewidth]{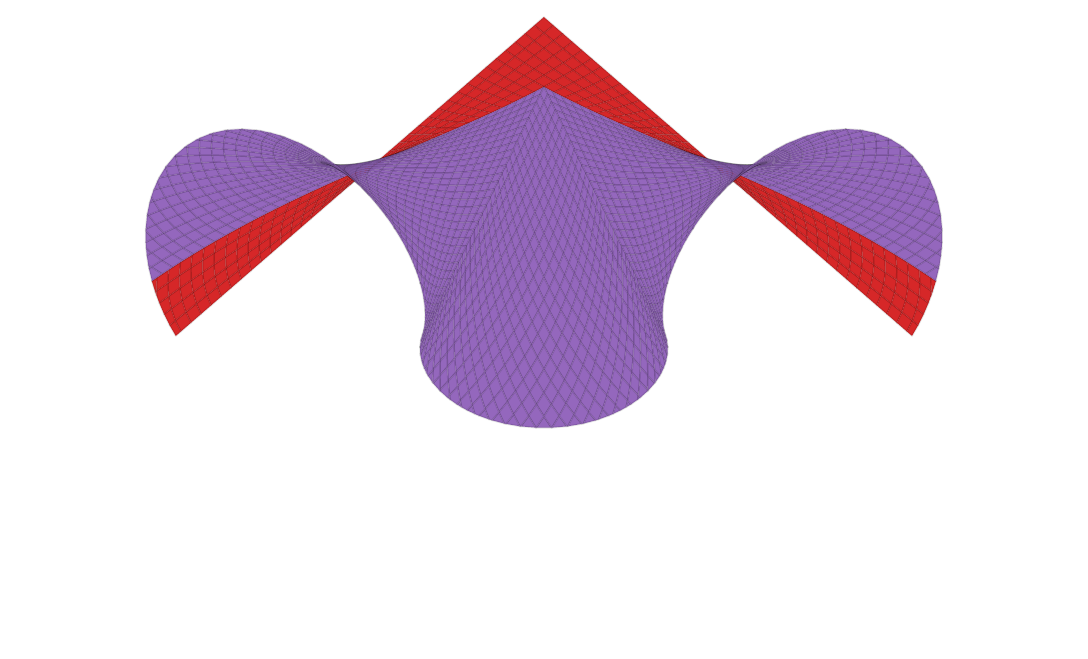}
                \caption{$r=2.25$}
                \label{fig:growth3}
        \end{subfigure}
        
        \begin{subfigure}[b]{0.8\linewidth}
                \includegraphics[trim={1.cm, 0cm, 0.65cm, 6cm}, clip, width=\linewidth]{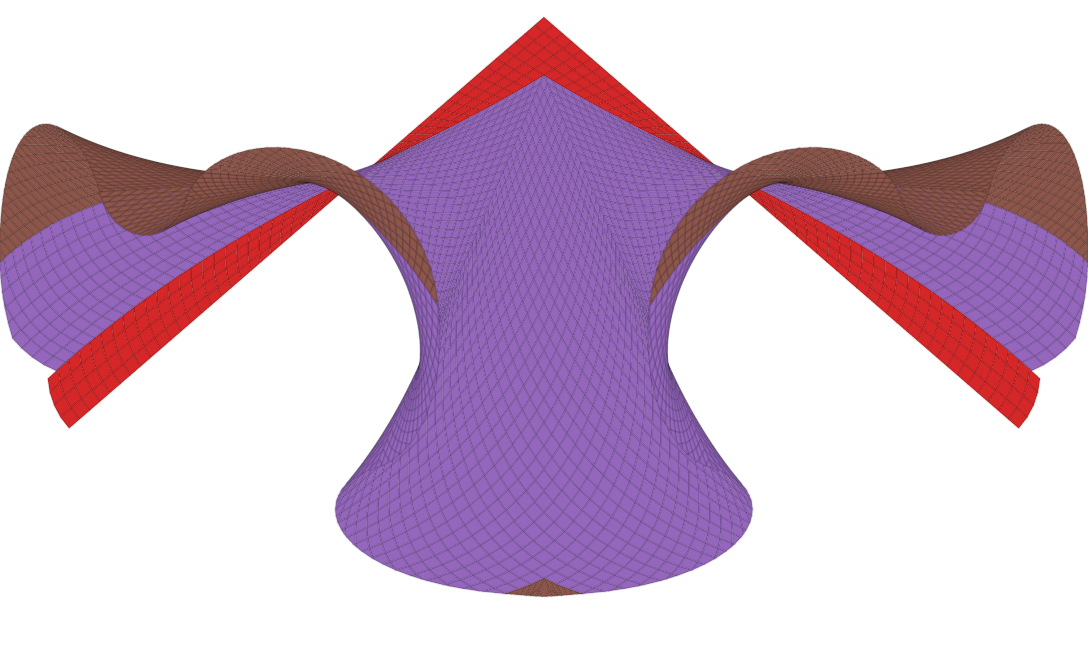}
                \caption{$r=2.90$}
                \label{fig:growth5}
        \end{subfigure}       
        \caption{Frames from a growing surface showing the process by which branch points enter from the boundary in a continuous way. The surfaces are colored by branch point generation.}
        \label{fig:growth}
\end{figure}

 One possibility is to have a single branch point at the origin with an index $m(t)$ that is increasing in time. Since $m(t)$ is discrete, it will be necessarily discontinuous in time; which would lead to global, discontinuous transitions of the geometry on increments of $m \to m+1$.  Another possibility is for branch points to enter the domain through the boundary as needed, and then move towards the origin continuously in time, allowing for continuous changes to geometry, rather than global shape transitions.
This effect is demonstrated by numerical experiments in the non small-slope setting and is illustrated in Fig.~\ref{fig:growth}. Sectors are colored by branch generation, i.e. the the color changes each time a new generation of branch points enters the surface.

Branch points entering through the boundary represent local shape deformations that are effected by changes in the asymptotic skeleton. With a fixed skeleton, $S$, changing the embedding $S \to \Omega \subset \mathbb{H}^2$ allows for movement of the existing branch points within the material, and changes the (Eulerian) morphology of the sheet. Branch points are not material (Lagrangian) entities. There are a large number of ``floppy" bending modes from varying the embedding $S \to \Omega$. These two mechanisms, namely branch points entering through the boundary and branch points moving relative to the material, are illustrated in Fig.~\ref{fig:growth}. We demonstrate two generations of branch points entering through the boundary and migrating towards the center as the sheet grows. 

\subsection{Shape control through distributed branch points} \label{sec:shapecontrol}
The technological applications of thin and flexible elastic sheets are growing more rapidly than ever with the advent of elastomeric and hydrogel thin-films \cite{kim2012designing} and soft robotics \cite{robotics}. To utilize these new technologies, understanding how one may control the shape of these structures becomes imperative, and hyperbolic geometries represent a significant challenge because they are intrinsically floppy. 

\begin{figure}[hbpt]
        \begin{subfigure}[b]{0.48\linewidth}
                \centering
                \includegraphics[width=.8\linewidth]{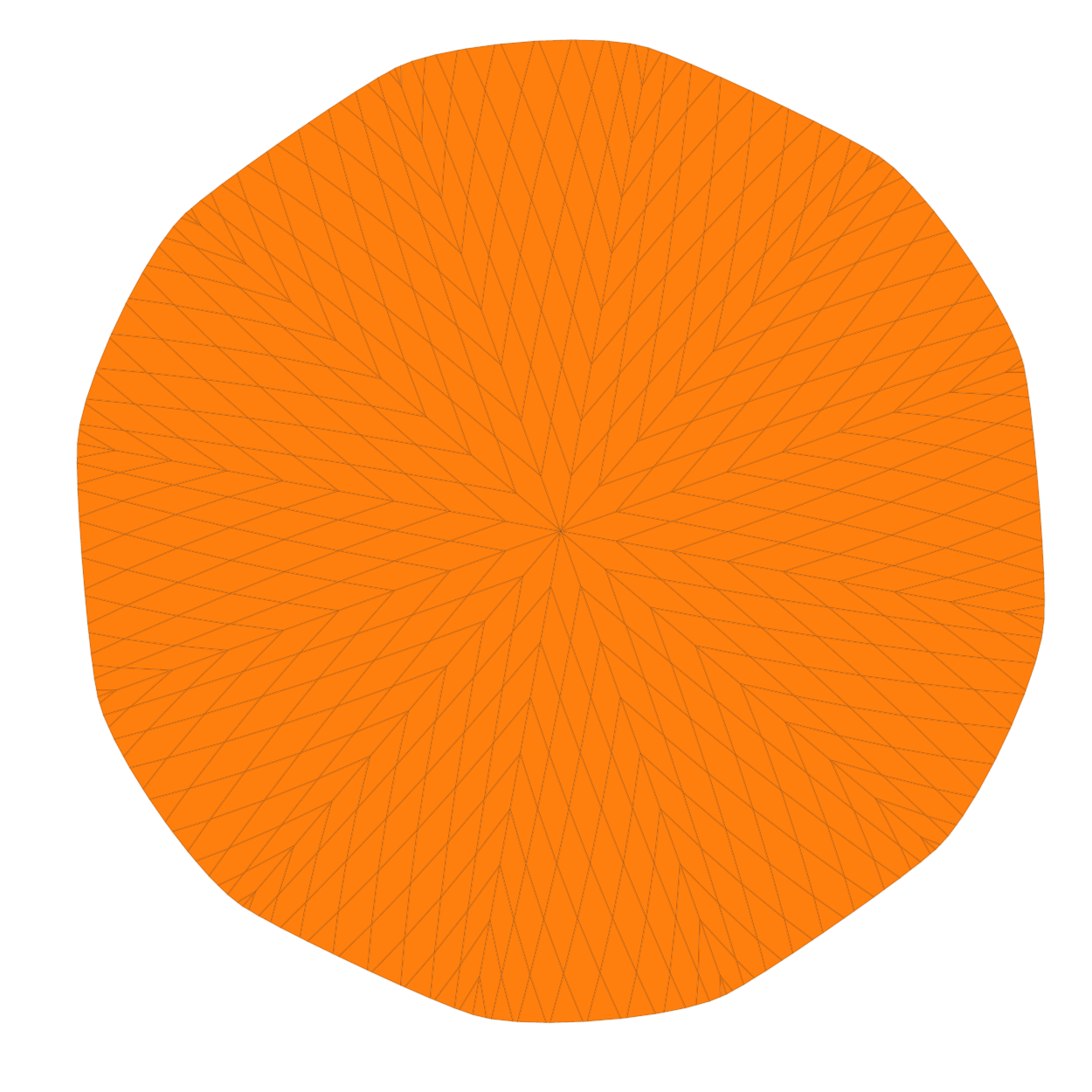}
                \caption{$r^*=0$}
                \label{fig:growth0}
        \end{subfigure}%
         \begin{subfigure}[b]{0.48\linewidth}
                \centering
                \includegraphics[width=.8\linewidth]{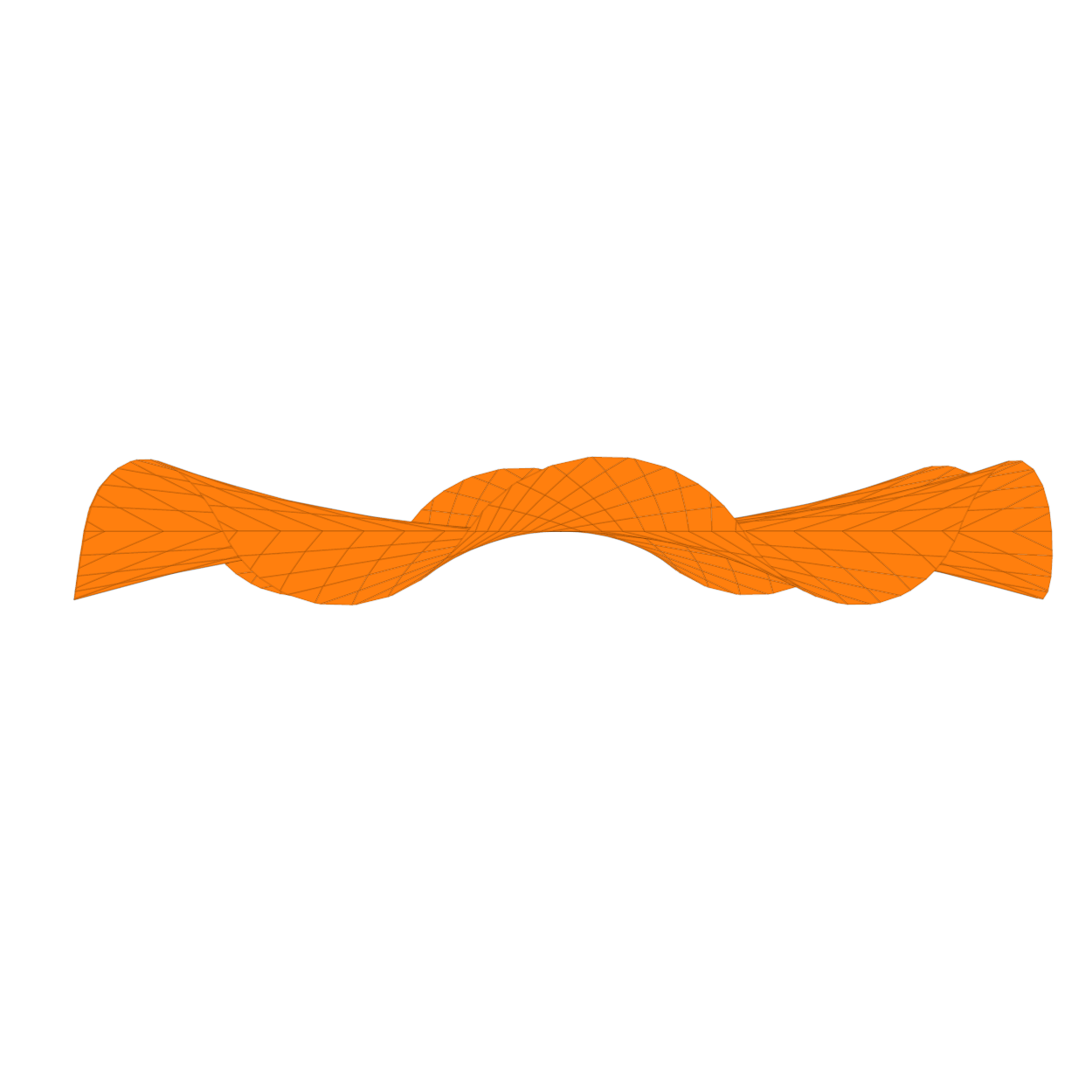}
                \caption{$r^*=0$}
                \label{fig:growthside0}
        \end{subfigure}%

        \begin{subfigure}[b]{0.48\linewidth}
                \centering
                \includegraphics[width=.8\linewidth]{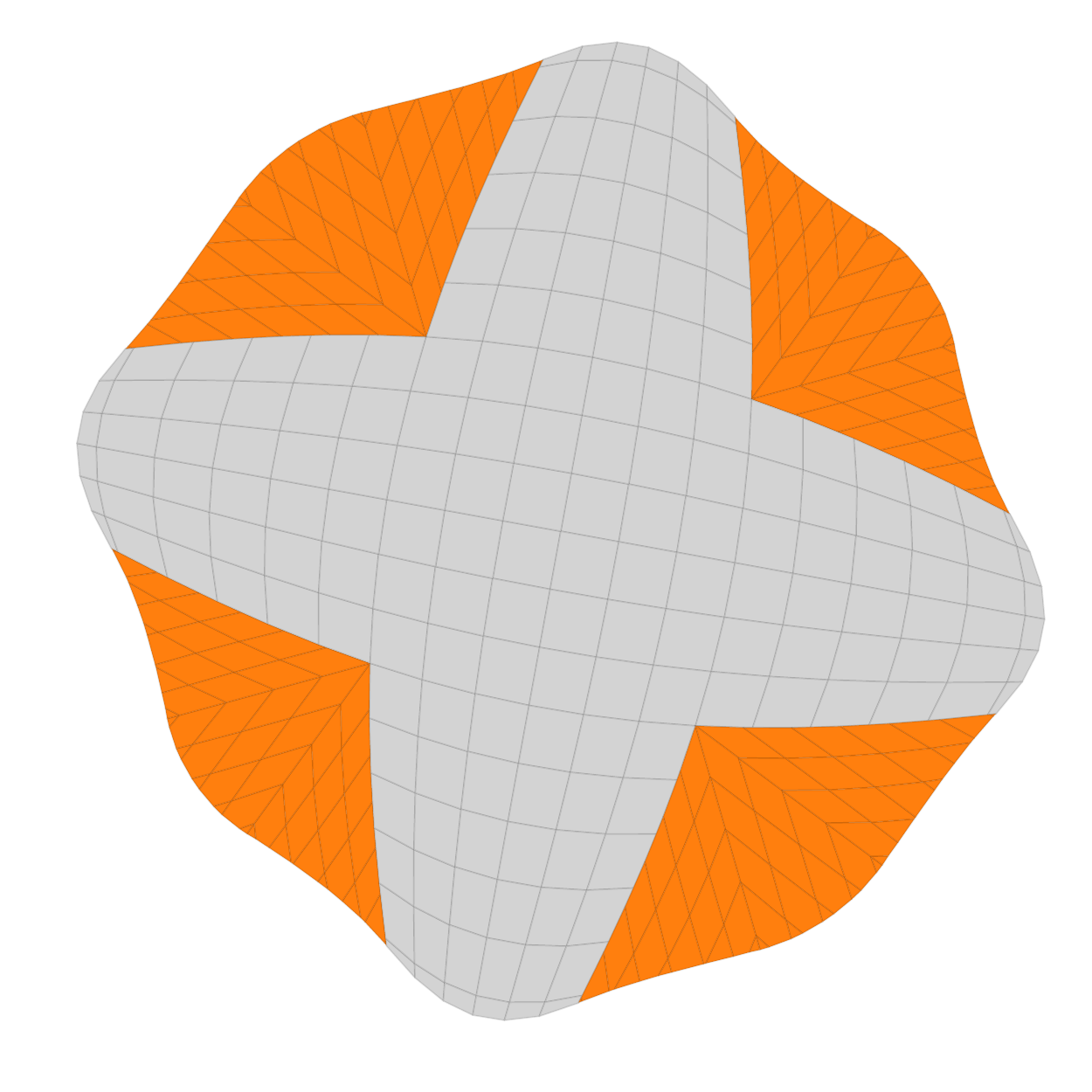}
                \caption{$r^*=0.5$}
                \label{fig:growth1}
        \end{subfigure}%
         \begin{subfigure}[b]{0.48\linewidth}
                \centering
                \includegraphics[width=.8\linewidth]{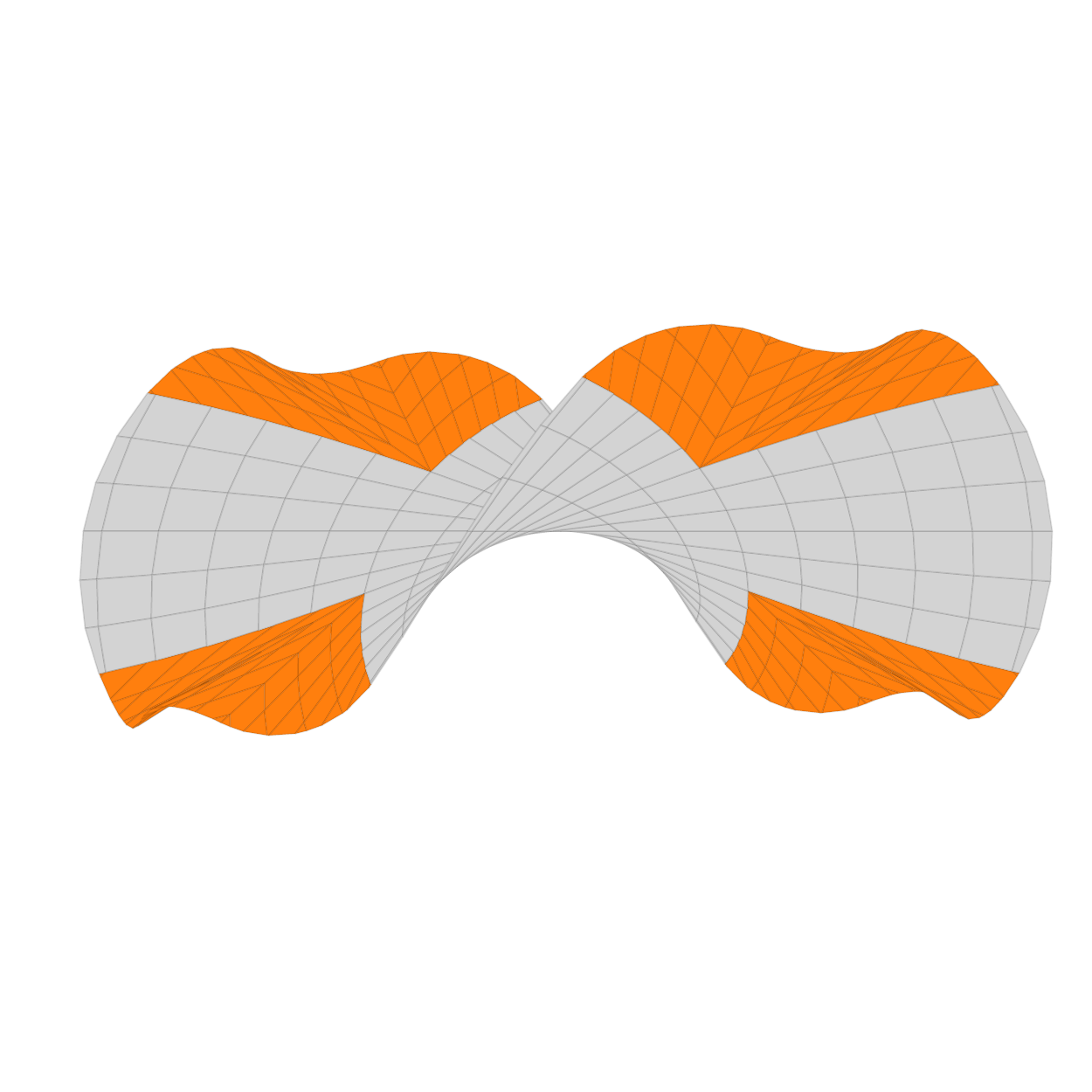}
                \caption{$r^*=0.5$}
                \label{fig:growthside1}
        \end{subfigure}
        
        \begin{subfigure}[b]{0.48\linewidth}
                \centering
                \includegraphics[width=.8\linewidth]{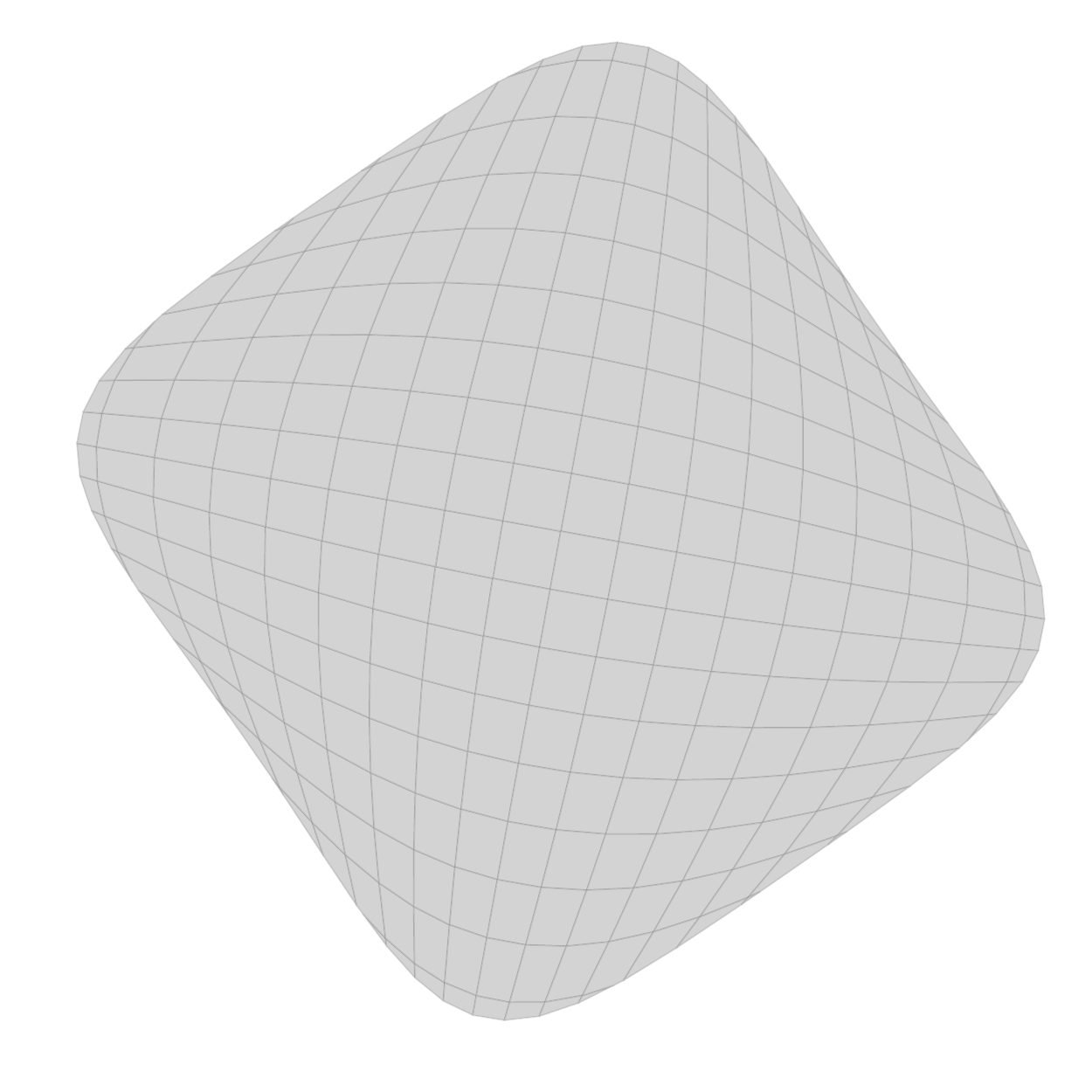}
                \caption{$r^* =1$.}
                \label{fig:growth2}
        \end{subfigure}%
        \begin{subfigure}[b]{0.48\linewidth}
                \centering
                \includegraphics[width=.8\linewidth]{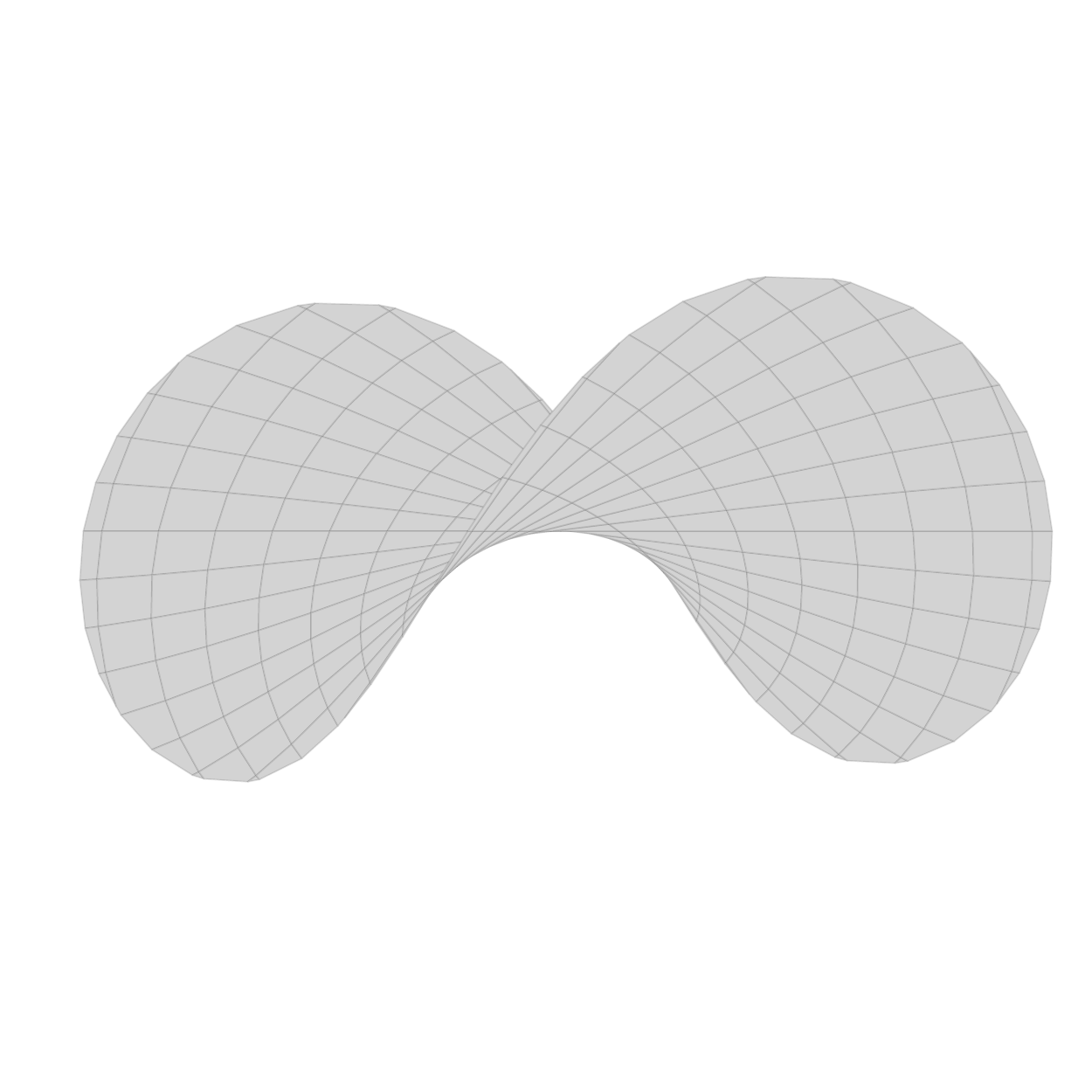}
                \caption{$r^* =1$.}
                \label{fig:growthside2}
        \end{subfigure}        
        \caption{A sequence of $R=1$ hyperbolic surfaces with varying branch-point defect locations, in fractional radius, along the diagonal. The  images on the left (a,c,e) show the surfaces from above. The colored regions are the 2\textsuperscript{nd} generation sectors. The images on the right (b,d,f) are the onside view. $r^*=1$ in (e,f) corresponds to no branch points on the sheet. The physical scales, in particular the vertical scales are the same for all the illustrations.}
        \label{fig:shapeprocedure}
\end{figure}

Motivated by the dynamics of growing sheets that were considered in the previous section, we now propose a novel idea for shape control - {\em branch point engineering}, which we illustrate by means of an example. Consider a hyperbolic disk $B = B_1$ of radius $R=1$ and intrinsic curvature $K = -1$ in the setting of the full (non small-slope) geometry. This disk can be isometrically embedded in $\mathbb{R}^3$ as a subset of the Amsler surface \cite{amsler1955surfaces,gemmer2011shape}. Imagine that we can engineer a branch point at a desired location along the diagonal $(u^*, u^*)$ by trisecting the local angle between the asymptotic directions. Since the original surface has four sectors, this will correspond to introducing four symmetrically placed branch points. This process is illuminated using Fig. \ref{fig:shapeprocedure} and is the ``full geometry" analog \cite{Shearman2021Distributed} of the ``small-slopes surgery" illustrated in Fig.~\ref{fig:SelfSimilar}.

\begin{figure}[ht]
    \center
    \includegraphics[width=.95\linewidth]{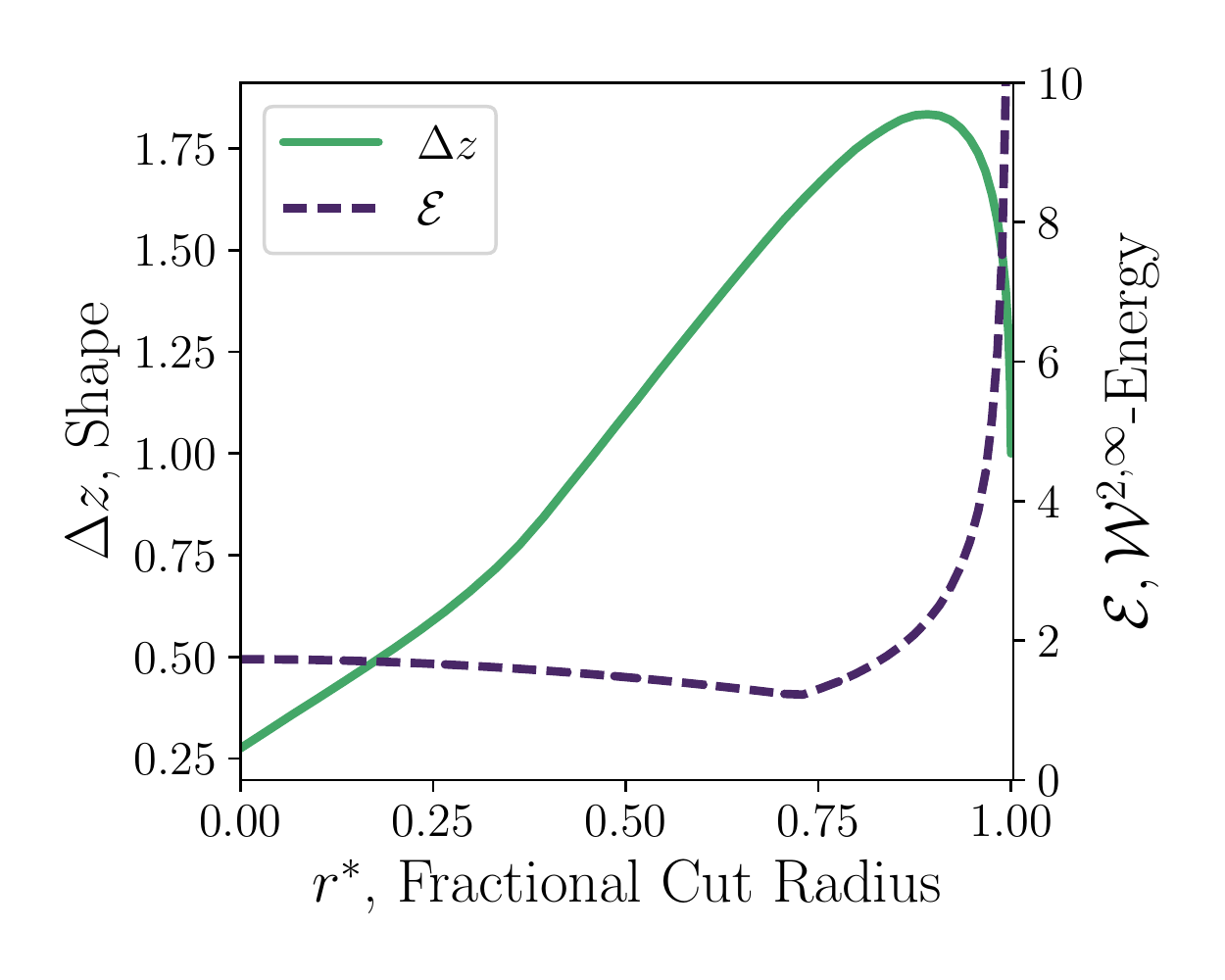}
    \caption{Total height, $\Delta z$ as a function of fractional branch point location, $r^*~=~\frac{r(u^*, u^*)}{r_{\max}}$ (solid) and the corresponding elastic energy (dashed)}
    \label{fig:shapecontrol}
\end{figure}

We have one control parameter, the geodesic radius $r^* = r(u^*,v^*)$ at the location of the branch point. For the resulting surface the total variation in the vertical coordinate, $\Delta z = z_{\max} - z_{\min}$ is a proxy for ``shape.'' Corresponding to each $r^* \in [0,1]$ we also compute the maximum curvature (the ``$W^{2,\infty}$ bending energy") $\kappa_{\max}$ of the surface. The dependence of $\Delta z$ and $\kappa_{\max} \equiv \mathcal{E}_\infty$ on $r^*$ are shown in Fig. \ref{fig:shapecontrol}.

The bending content $\mathcal{E}_2= \int_B (\kappa_1^2+\kappa_2^2)$ can be estimated in terms of the maximum curvature by $\mathcal{E}_2 \lesssim 2 A \kappa_{\max}^2$
where $A \sim R^2$ is the  area of the sheet of radius $R$. We can estimate the (physical) force resisting vertical compression by
\begin{align}
 \label{eq:force}
    F  \approx & - \frac{Y R^2}{24} \left(\frac{t}{R}\right)^3 \frac{1}{R^{-1}}\frac{d \mathcal{E}_2}{d (\Delta z)} \\
     \simeq & -  \frac{Y t^3 \kappa_{\max}}{6} \left(\frac{d (\Delta z)}{d r^*}\right)^{-1} \left(\frac{d \kappa_{\max}}{d r^*}\right), \nonumber 
\end{align}
where $Y$ is the elastic modulus of the material and $t$ is the thickness of the sheet. The expression follows from recognizing that $YR^2$ is a force scale, $\mathcal{E}_2$ and $R^{-1} \Delta z$ are both dimensionless, and the physical elastic energy scales as $t^3$ with the thickness of the sheet.  

Of course, the right hand side of Eq.~\eqref{eq:force} is only an estimate of the true force. Nonetheless, it is a useful expression for describing qualitative behavior. There is a wide plateau, between $r^* = 0$ and $r^* \approx 0.75$ where the energy changes very little despite an almost seven fold change in $\Delta z$. The force in this regime resists compression but it is small, implying that the surface is very floppy. Between about $r^* = 0.75$ and $r^* = 0.9$ the force no longer resists compression, so the sheet is mechanically unstable and can act as a switch spontaneously generating forces on the scale $Y t^3/R$. Finally there is an outer region $r^* \gtrsim 0.9$ where the sheet acts as a ``stiffer" spring in comparison to the regime $r^* \lesssim 0.75$. Also, the branch points couple differently with the vertical compression -- for $r^* \lesssim 0.75$ vertical compression pushes the branch points inwards, while for $r^* > 0.9$ it pushes them outwards.

\subsection{Compressed gel experiments}\label{sec:expts}
\begin{figure}[ht]
    \center
    \includegraphics[width=\linewidth]{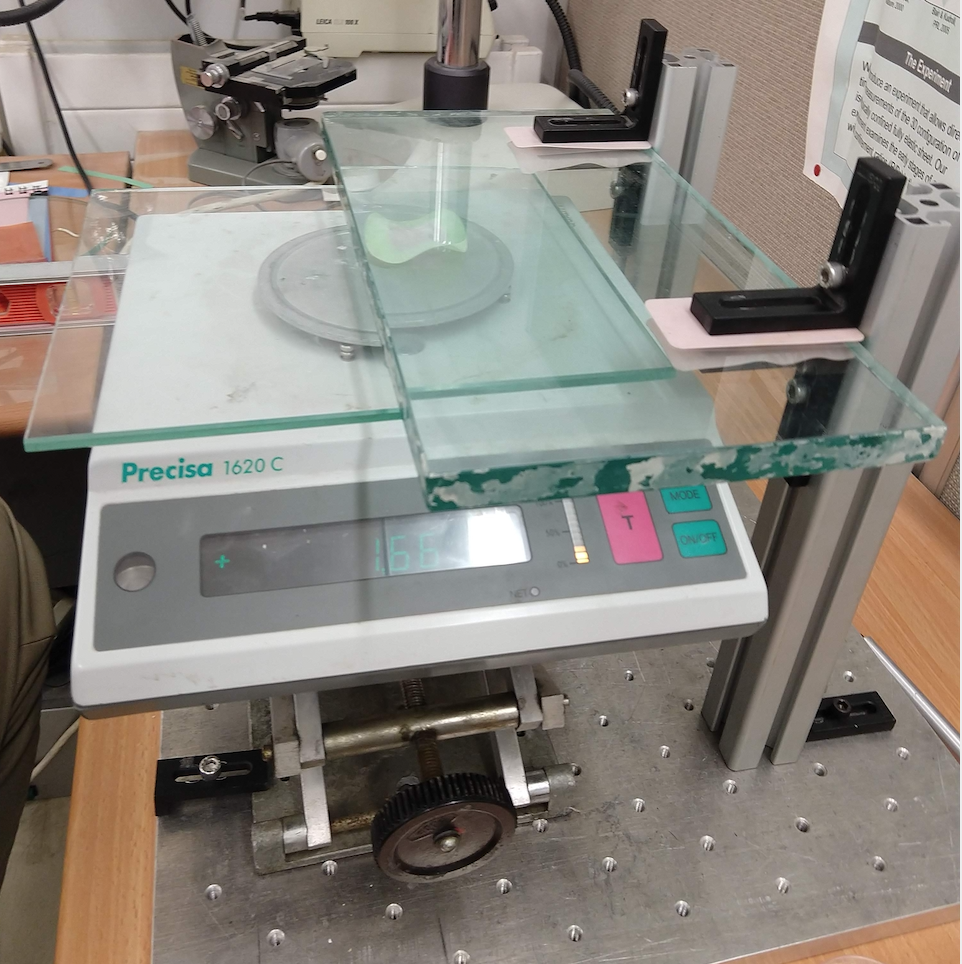}
    \caption{The experimental setup.}
    \label{fig:expt}
\end{figure}

While the mathematics of thin hyperbolic sheets and the role of branched $\mathcal{C}^{1,1}$ isometries form a compelling theory for the geometry and mechanics of slender elastic structures, it is still essential to connect these theoretical models with actual physical experiments. To this end, we performed an experiment to directly measure the net force in compressing a hyperbolic polymer gel in between two glass plates and observe the attendant morphological changes. The experimental setup is shown in Fig.~\ref{fig:expt} and consists of a stationary glass plate, leveled to be horizontal and clamped to a frame, and a second glass plate, also leveled to be horizontal set on a weighing scale, which is on a platform that can be raised and lowered. Between the plates is a  hyperbolic polymer gel sample that was cast in polyvinylsiloxane (PVS - Zhermack Elite Double 32) following the protocols in \cite{Pezzulla2015Morphing}.

As the platform was raised, the gel developed additional wrinkles in sudden transitions accompanied by a sharp decrease in the force against the plates. Typical force-displacement data are shown in Fig. \ref{F:experimentalData}. Although the force curves and wrinkle transitions were not exactly the same as in the forward direction due to hysteresis and friction, repeating the experiment backwards, i.e., increasing the displacement between the plates, qualitatively reproduced the same observations in reverse: The hyperbolic gel transitioned from a plane-stress to a highly wrinkled state, before losing wrinkles one at a time, and the force trend was retraced backwards.

Rather than model the precise geometry of the gel, our approach was to deduce the effective mechanical/geometric properties of the sample entirely from the force-displacement curves. This approach is along the lines of the original experiments in Pezzulla et al \cite{Pezzulla2015Morphing} and can be viewed as a verification that a simple model, treating the gel as an elastic surface with a constant target curvature with a geometry that is in the small-slopes regime, is entirely adequate for the purposes of modeling the forces and the shape transitions in this sample.

\subsubsection{Experimental data}\label{S:experimentalData}

\begin{figure}[htbp]
\begin{subfigure}[t]{0.95\linewidth}
\includegraphics[width=\textwidth]{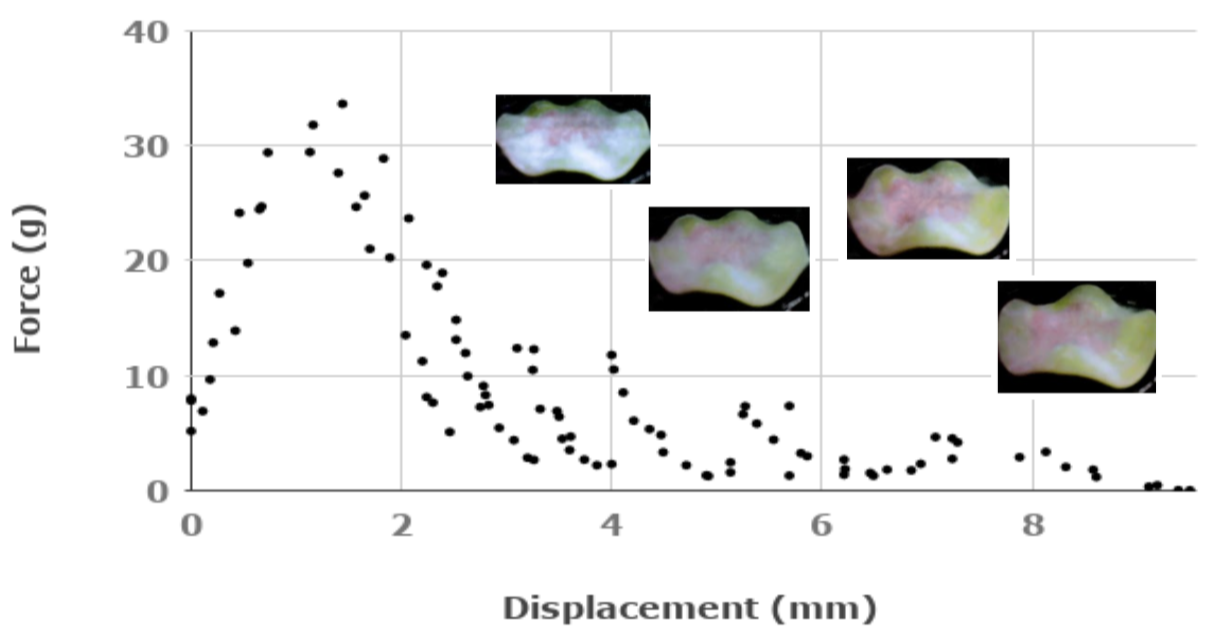}
\end{subfigure}\\
\begin{subfigure}[t]{0.95\linewidth}
\includegraphics[width=\textwidth]{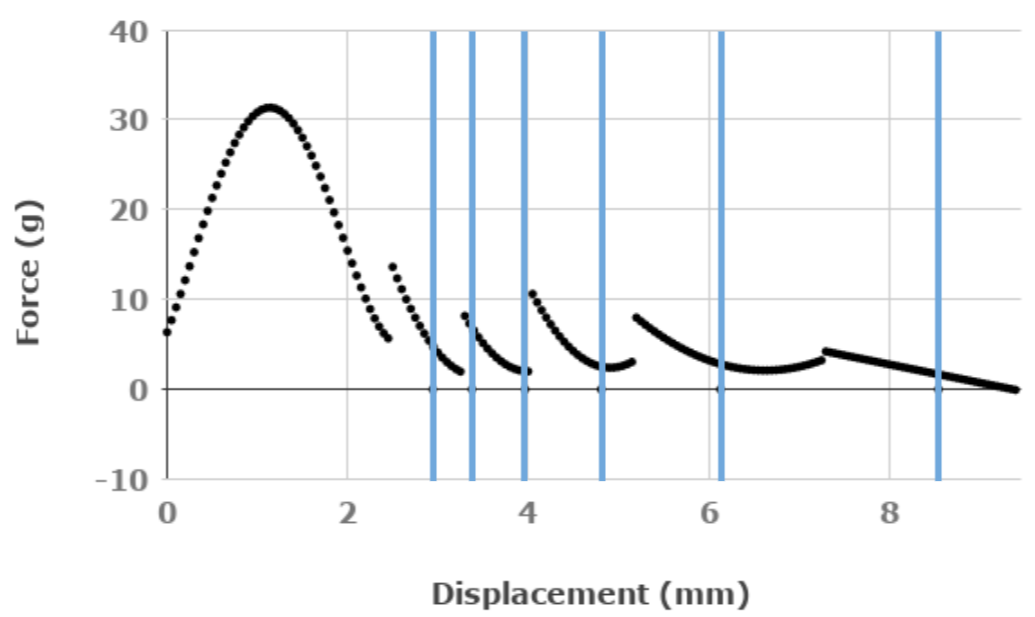}
\end{subfigure}
\begin{subfigure}[t]{0.95\linewidth}
\includegraphics[width=\textwidth]{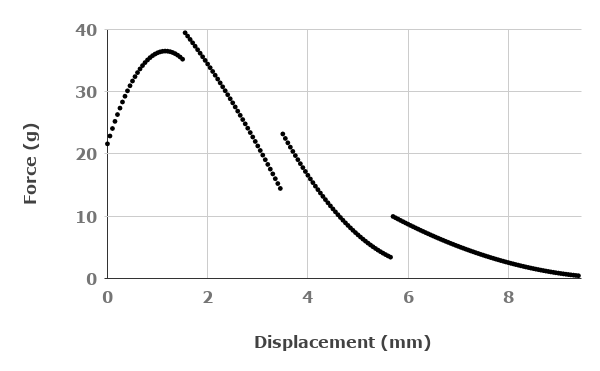}
\end{subfigure}
\caption{Net vertical force (in gram-force) exerted by a compressed sheet as a function of displacement (in millimeters) of the glass plates. Data for the top two plots were obtained by increasing the displacement between the glass plates, starting from when the gel is in a plane-stress state. The top plot consists of raw data with photos of the hyperbolic gel morphology. The middle plot shows polynomial fits to the data. The vertical blue lines delineate predictions for the displacement at the minima of the curves for each wrinkle regime. The bottom plot shows polynomial fits to data obtained by decreasing the displacement between the glass plates, i.e., progressively compressing the gel until it reaches a plane-stress state.}
\label{F:experimentalData}
\end{figure}

The plots in Fig. \ref{F:experimentalData} show the force-displacement curve obtained when increasing the displacement between the glass plates, starting from a position with the gel in a plane-stress state. As the plates are separated, the gel first takes on eight wrinkles until wrinkles are successively shed one at a time leaving only  three when the plates are the furthest apart. On the bottom plot in Fig. \ref{F:experimentalData}, each separate polynomial curve represents a different wrinkle regime. The leftmost curve corresponds to when the gel is at a plane-stress state before transitioning to eight wrinkles. Each successive polynomial curve to the right sheds one wrinkle. The rightmost nearly-linear curve corresponds to three wrinkles. Polynomial regression was used to model the data over each wrinkle regime. A second-degree polynomial was used to fit the data for all wrinkle regimes, except for the leftmost curve corresponding to the plane-stress and eight-wrinkle regime which is of fourth-degree.

In the force-versus-displacement data, the wrinkles in the gel are the most symmetric at the minima of the curves for each respective wrinkle regime. The leftmost portion of the curves correspond to compressed wrinkles with flattened tops or bottoms. The rightmost portion of the curves, particularly when the force goes up slightly, corresponds to when the wrinkles are asymmetric as the top or bottom of a wrinkle gradually and increasingly separates from the glass plate until two adjacent wrinkles become one (larger) wrinkle.

\subsubsection{Predictions of wrinkle transitions based on \texorpdfstring{$\mathcal{C}^{1,1}$}{C1,1} isometries}
The blue vertical lines in the bottom plot of Fig. \ref{F:experimentalData} are analytical predictions for the displacement at the minima of the curves for each wrinkle regime. The predictions are especially accurate for the five through seven-wrinkle curves but less so for the others. Some discrepancies are expected since the predictions are based on computing the out-of-plane displacement for a small-slopes approximation for the immersion of a hyperbolic sheet with \emph{constant} negative Gauss curvature. In reality, the Gauss curvature is almost certainly not uniform over the entire gel. There are also significant inelastic and non-conservative effects---e.g., adhesion, friction, and hysteresis---affecting the wrinkle transitions that have not been modeled.

To obtain the predictions, we assume that the out-of-plane displacement for a single sector, with an angular extent $\theta$ and aligned symmetrically along the positive $x$-axis, is given by
\begin{equation*}
w(x,y)=\frac{\sqrt{|K|}}{2}\left(x^2\tan\frac{\theta}{2}-y^2\cot\frac{\theta}{2}\right),
\end{equation*}
which solves the Monge Amp\`ere equation $w_{xx}w_{yy}-w_{xy}^2=-|K|$, where $K$ is the constant intrinsic Gauss curvature for the gel.

Since the edges of this sector have zero out-of-plane displacement, the total vertical extent of the sector is given by the maximum out-of-plane displacement which, for a sheet of radius $R$, occurs at $(x,y)=(R,0)$ and is
\begin{equation*}
w(R,0)=\frac{\sqrt{|K|}}{2}R^2\tan\frac{\theta}{2}.
\end{equation*}

Assuming a symmetric $\mathcal{C}^{1,1}$ piecewise solution for the out-of-plane displacement, $w$, which glues together upward- and downward-curving sectors of equal angular extent over a disc, the angular extent is $\theta=\pi/n$ for a single sector when there are $n$ (symmetric) wrinkles. Then, the total vertical extent $\bar{w}(n)$ of the out-of-plane displacement of the gel when there are $n$ wrinkles is twice the vertical extent of a single sector,
\begin{equation*}\label{E:verticalExtent}
\bar{w}(n)=2\cdot\frac{\sqrt{|K|}}{2}R^2\tan\frac{\pi}{2n}=\sqrt{|K|}R^2\tan\frac{\pi}{2n}.
\end{equation*}

Note that $\bar{w}(n)$ predicts the vertical extent when there are $n$ wrinkles that are as symmetric as possible and do not have tops or bottoms flattened by the glass plates. This corresponds to the displacement at the minima (or the rightmost point) of the curves for each respective wrinkle regime, as described at the end of Section \ref{S:experimentalData}. 

Since the intrinsic Gauss curvature $K$ of the hyperbolic gel is unknown, we treat $\sqrt{|K|}$ as a free parameter that is chosen so that the function $\bar{w}(n)$ best predicts the displacement at the minima of the force-versus-displacement curves for $n=3,...,8$. With the radius of the gel measured to be $R\approx23.67\,\SI{}{\milli\meter}$, we compute the various values for the free parameter $\sqrt{|K|}$, shown in Table \ref{T:freeParameter}, such that $\bar{w}(n)$ matches the displacement at the minima of the curves for $n=3,...,8$, i.e., $\bar{w}(3)=9.35$, $\bar{w}(4)=6.65$, $\bar{w}(5)=4.9$, $\bar{w}(6)=4$, $\bar{w}(7)=3.25$, and $\bar{w}(8)=2.45$.

\begin{table}[htbp]
\caption{\textbf{Free parameter values.} Values for the free parameter $\sqrt{|K|}$ (in $\SI{}{\per\milli\meter}$) so that $\bar{w}(n)$ matches the displacement at the minima of the force-versus-displacement curves for the number of wrinkles $n=3,...,8$. }
\centering
\begin{tabular}{c|c} 
	\toprule
	$n$&$\sqrt{|K|}$\\
	\midrule
	3&0.0289\\ 
	4&0.0287\\
	5&0.0269\\
	6&0.0266\\
	7&0.0254\\
	8&0.0220\\
	\bottomrule
\end{tabular}
\label{T:freeParameter}
\end{table}

It is notable that these values turn out to be relatively close to each other, as shown in Table \ref{T:freeParameter}. This indicates that there is indeed a value for the free parameter $\sqrt{|K|}$ such that even this simple theory, based on the vertical extent of symmetric $\mathcal{C}^{1,1}$ piecewise solutions for the out-of-plane displacement, satisfactorily predicts the wrinkle transitions in the experiment.

Using the average of these values for the free parameter, i.e., $\sqrt{|K|}\approx0.0264\,\SI{}{\per\milli\meter}$, we compute $\bar{w}(3),...,\bar{w}(8)$ and compare with their experimental values in Table \ref{T:compareWithExperiment}. As mentioned before, the vertical lines in the bottom plot in Fig. \ref{F:experimentalData} also mark the predicted vertical displacements $\bar{w}(3),...,\bar{w}(8)$ for the minima of the curves for each wrinkle regime. The predictions are within a relative error of 10\%, except for $n=8$ which has a relative error of 20\% which may be attributed to the prevalence of stretching which violates the isometry assumption of our model (Table \ref{T:compareWithExperiment}).

\begin{table}[htbp]
\caption{\textbf{Comparison between prediction and experimental data.} Predicted and experimental values for the displacement (in $\SI{}{\milli\meter}$) at the minima of the force-versus-displacement curves for number of wrinkles $n=3,...,8$. The relative error is within 10\%, except for $n=8$.}
\centering
\begin{tabular}{c|c|c|c} 
\toprule
$n$&$\bar{w}(n)$&Experimental&Relative Error (\%)\\
\midrule
3&8.55&9.35&8.56\\ 
4&6.13&6.65&7.82\\
5&4.81&4.9&1.84\\
6&3.97&4&0.75\\
7&3.38&3.25&4\\
8&2.94&2.45&20\\
\bottomrule
\end{tabular}
\label{T:compareWithExperiment}
\end{table}

\subsection{Weak forces: Scaling laws and smoothness}\label{S:WeakForcesScaling}

We now investigate the floppiness of thin hyperbolic sheets by considering the effects of a distributed `weak' force, here modeled by gravity, on the shape of the sheet. In this setting, stretching dominates all the other forces in the problem, and it therefore suffices to minimize the sum of bending and gravitational energies over all isometric conformations of the sheet. Note that the isometry constraint Eq.~\eqref{eq:mng_amp_cnst} and the bending content only depend on the out-of-plane displacement $w(x,y)$. We thus need to minimize, over the choice of $w(x,y)$, the following constrained energy functional:
\begin{align}
\frac{\mathcal{E}[F]}{R^4} & =\frac{\lambda}{4R^2}\int_D\left(w_{xx}+w_{yy}\right)^2\,dxdy +\int_Dw\,dxdy, \nonumber\\
w & \geq 0, \quad \mathrm{det} (D^2w)  = -1,
\label{E:dimensionlessBendingGravity}
\end{align}
where $\lambda$ is the ratio of bending modulus to the gravitational force density $\rho g$. With the assumption that both of these parameters scale as $t^2$, it follows that $\lambda$ is independent of $t$ and the above energy, along with the isometry constraint, therefore describes the $t \to 0$ limit.

As we discussed above, in the vicinity of a branch point, which we take without loss of generality to be at the origin, a piecewise quadratic solution to $\mathrm{det}(D^2 w) = -1$ on two adjacent sectors is given by 
\begin{align}\label{E:mongeAmpereSolution}
w(x,y)=& ax+by+c \\
&+\begin{cases}y(x-y\cot\theta_+),&0\le\theta\le\theta_+\\y(x+y\cot\theta_-),&-\theta_-\le\theta\le0 \end{cases}, \nonumber
\end{align}
where $0<\theta_{\pm}<\pi$; $a$, $b$, and $c$ are constants; and $(x,y)$ and $(r,\theta)$ denote Cartesian and polar coordinates on $\mathbb{R}^2$. Here, the surface $w(x,y)$ has a continuous tangent plane along the ray ${\theta=0}$ where the two pieces are attached and, hence, the glued surface has finite bending energy. 

The sector $0 \leq \theta \leq \theta_+$ is an {\em upward-curving sector} since the surface here is above the tangent plane to the origin $z = ax+by+c$, and correspondingly the section $\theta_- \leq \theta \leq 0$ is a {\em downward-curving sector}. The angular ratio for this pair of adjacent sectors is defined to be $\theta_+/\theta_-$.

Branch points are characterized by (i) the degree of the branch point, which is defined as half of the total number of incident sectors, and (ii) the angular extent of the upward- and downward-curving sectors, adding up to $2 \pi$. 

\begin{figure}[htbp]
\includegraphics[width= \linewidth]{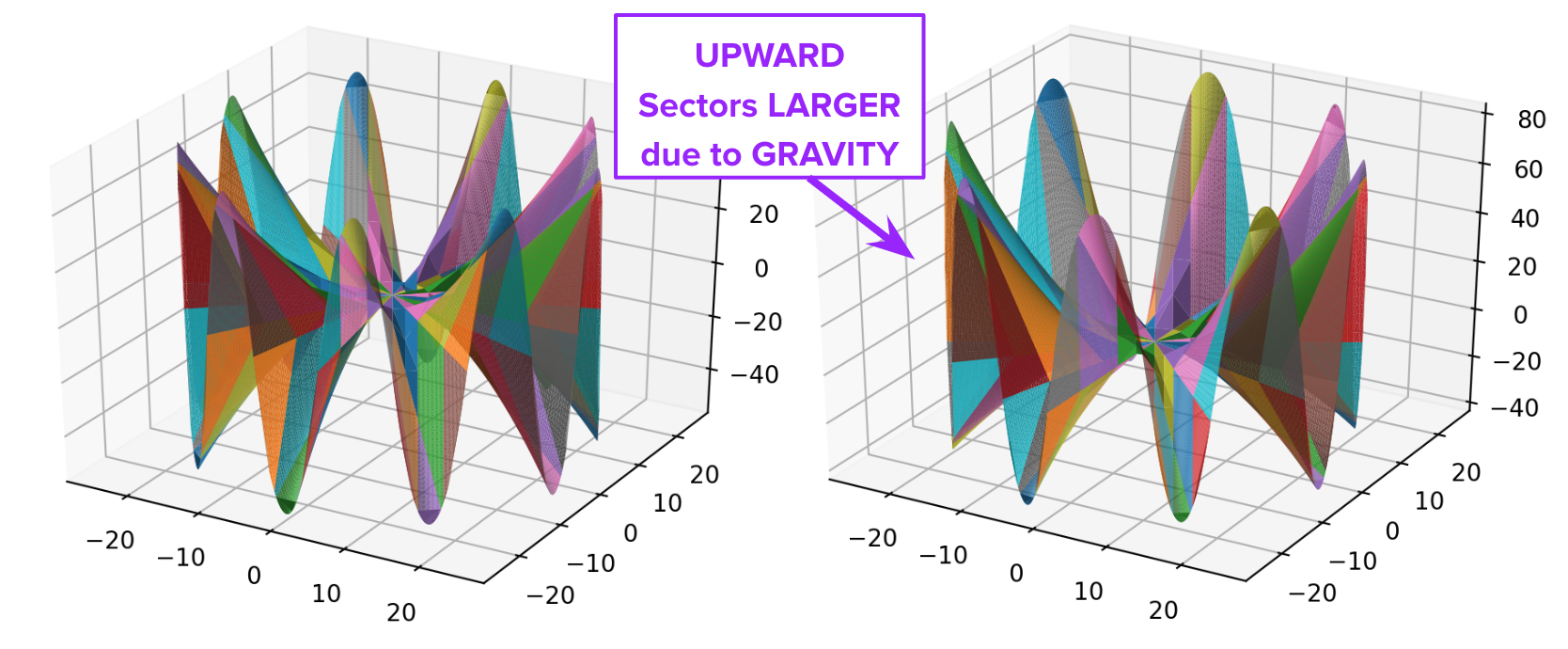}
\caption{Optimal angular ratio between upward- and downard-curving sectors: The center of mass of the surface is lowered by making the downward-curving sectors span a smaller angular extent than the upward-curving sectors.}
\label{F:upDownOptimization}
\end{figure}

Energetically, the preferred geometries from a pure-bending and pure-gravity perspective are in direct opposition. Bending prefers lower-degree branch points and equal angular extents, while gravity prefers higher-degree branch points and asymmetric angular extents that lower the center of mass. Depending on whether the system is in a bending- or gravity-dominant regime, we observe more of either qualities in the energy-minimizing geometry (Fig. \ref{F:upDownOptimization}).

In the gravity-dominant limit $\lambda \ll 1$, the optimal angular ratio between the upward- and downward-curving sectors for $\mathcal{C}^{1,1}$ surfaces is 2:1. This is also the case with $\mathcal{C}^2$ surfaces with straight asymptotic lines. This is confirmed analytically and numerically in a forthcoming paper \cite{Yamamoto2021Role}.

\begin{figure}[htbp]
\includegraphics[width= \linewidth]{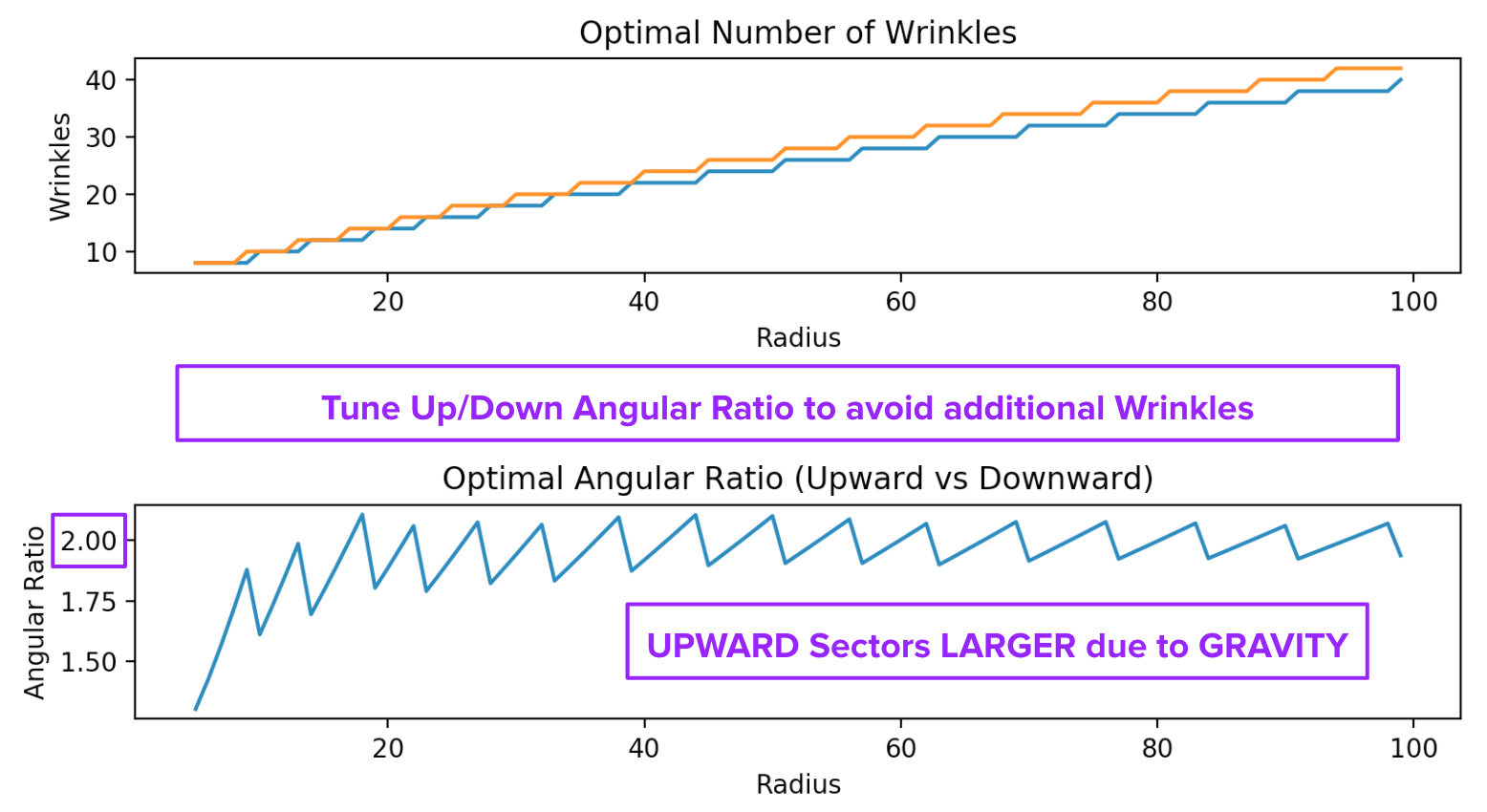}
\caption{Optimal number of wrinkles and angular ratio as a function of radius. While the orange curve is for $\mathcal{C}^{1,1}$ isometries with upward- and downward-curving sectors restricted to be of equal angular extent, the blue curve is for surfaces that allow these sector angles to be different. As the radius is increased, the optimal angular ratio between the upward- and downward-curving sectors approaches 2.
}
\label{F:singleBranchPointOptimizedFeatures}
\end{figure}

At smooth ($C^2$) points, there are 4 sectors, so a ``non-branch" point has degree 2. In contrast, a branch point has degree $\geq 3$ and we define the excess degree by subtracting 2 from the degree of a branch point.   The excess degrees of the branch points add, so that the total number of wrinkles in the sheet, which is equal to the number of sectors that intersect the boundary, is given by 
$$
N_{wrinkles} = 4 + 2 \sum_i (d_i -2)
$$
where the sum is over all the branch points on the surface and $d_i$ is the degree of the $i$-th branch point. For piecewise quadratic surfaces with a single branch point at the origin, the total number of wrinkles is the same at all radial locations and equals twice the degree of the origin. The scaling of the number of wrinkles for the (restricted) minimizers with equal up-down angles, as well as independent up-down angles is shown in Fig.~\ref{F:singleBranchPointOptimizedFeatures}. Although the scaling of the total number of wrinkles $N_{wrinkles}$ with respect to the radius is the same, allowing for an asymmetry between the upward- and downward-curving sectors slightly delays the introduction of a new wrinkle with increasing radius of the domain. This is illustrated in the top plot in Fig. \ref{F:singleBranchPointOptimizedFeatures}. The bottom plot shows that, before the creation of another wrinkle, the angular ratio between the upward- and downward-curving sectors gradually increases to approximately 2:1. It turns out that this is energetically more efficient until the radius becomes large enough that it is beneficial to create an additional wrinkle. Then, in the instant that a new wrinkle is created, the angular ratio lowers, making the upward- and downward-curving sectors more symmetric again.

\begin{figure}[htbp]
\includegraphics[width= \linewidth]{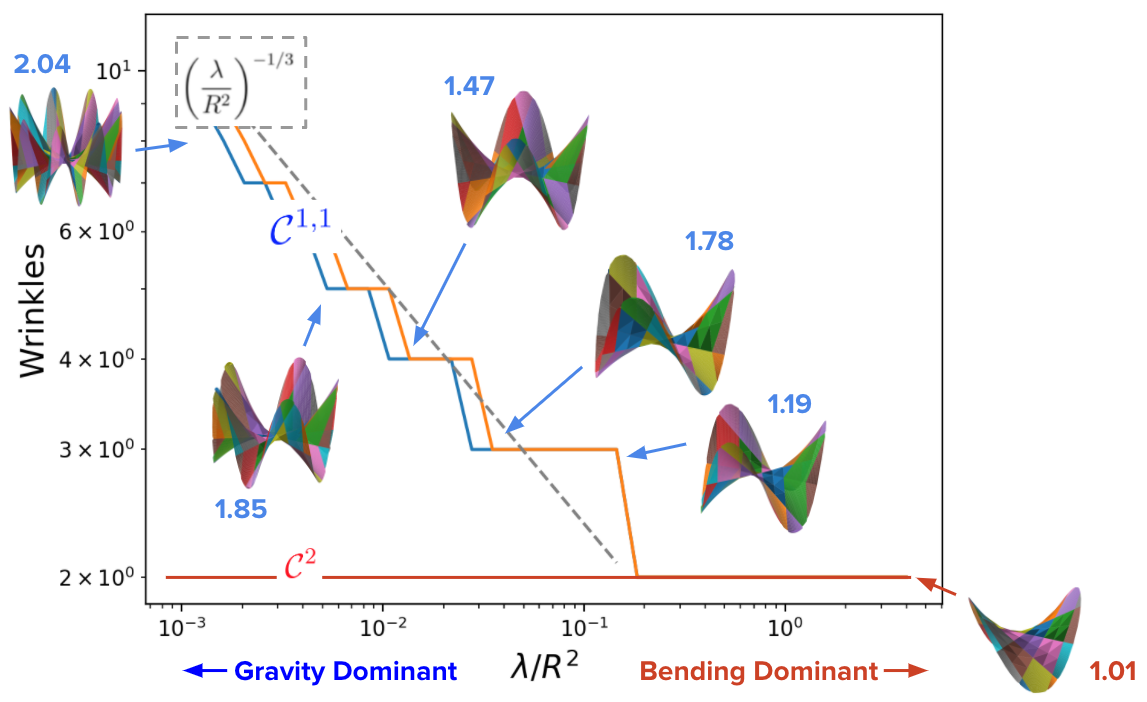}
\caption{Scaling of the number of wrinkles for static hyperbolic sheets: Branch points allow for wrinkles which lower the sheet's center of mass and, hence, gravitational potential energy. The red $\mathcal{C}^2$ lines is for a $\mathcal{C}^2$ surface with straight asymptotic lines. The orange and blue lines are for $\mathcal{C}^{1,1}$ isometries with a single branch point at the origin with upward- and downward-curving sectors of equal or different angular extent, respectively. The number by each image of a hyperbolic sheet is the angular ratio between the upward- and downward-curving sectors.}
\label{F:nonDimScalingWrinkles}
\end{figure}

Balancing the bending and gravity energies reveals a power-law relationship between the optimal number of wrinkles (corresponding to the lowest total energy) and the radius $R$, where the optimal number of wrinkles is proportional to $R^{2/3}$. Subsequently, it can also be shown that the energy scales as $R^{10/3}$. The number of wrinkles $n$ with respect to the dimensionless parameter $\lambda/R^2$ for $\mathcal{C}^{1,1}$ isometries scales as $n\sim\left(\lambda/R^2\right)^{-1/3}$. These scaling laws can be confirmed both numerically and analytically; see Fig. \ref{F:nonDimScalingWrinkles}.

Given the zero-stretching constraint given by Eq.\eqref{eq:mng_amp_cnst}, we have
\begin{equation*}
\kappa_\theta\kappa_r=-1,
\end{equation*}
where $\kappa_\theta$ and $\kappa_r$ are the principal curvatures in the azimuthal and radial directions, respectively. For $n$ wrinkles, i.e., pairs of upward and downward curving sectors, we have
\begin{equation*}
\kappa_\theta\sim\pm\frac{n}{\pi}\quad\text{and}\quad\kappa_r\sim\mp\frac{\pi}{n}.
\end{equation*}
With this constraint and assuming a $\mathcal{C}^{1,1}$ piecewise solution $w(x,y)$ to the zero-stretching constraint PDE given by Eq.\eqref{eq:mng_amp_cnst} given by odd extensions of Eq. \eqref{E:mongeAmpereSolution} over a unit disc $(x,y)\in D\subset\mathbb{R}^2$ with a horizontal tangent plane at the origin (i.e., $a=b=0$), we have that the vertical height $w_\text{max}-w_\text{min}$ of the solution is
\begin{equation*}
w_\text{max}-w_\text{min}\sim2r^2|\kappa_r|\sim2l^2|\kappa_\theta|\sim\frac{2\pi}{n},
\end{equation*}
where $l=\pi/n$ is the arclength of the $xy$-projection of a single upward or downward curving sector of the $\mathcal{C}^{1,1}$ solution over $D\subset\mathbb{R}^2$ and $r=1$ is its radius. Consequently, setting $w\sim w_\text{max}-w_\text{min}$ to approximate the scaled dimensionless energy functional given by Eq.\eqref{E:dimensionlessBendingGravity} we obtain
\begin{equation*}
\begin{aligned}
\frac{\mathcal{E}[F]}{R^4}&\approx\frac{\lambda}{R^2}\int_DH^2\,dxdy+\int_Dw\,dxdy
\\&\sim\frac{\lambda}{R^2}\int_D\left(\kappa_\theta+\kappa_r\right)^2\,dxdy 
\\& +\int_D\left(w_\text{max}-w_\text{min}\right)\,dxdy
\\&\sim\frac{\lambda}{R^2}\left(\pm\frac{n}{2\pi}\mp\frac{2\pi}{n}\right)^2+\frac{2\pi^2}{n}.
\end{aligned}
\end{equation*}
Therefore, assuming large $n$, we drop the $\mp2\pi/n$ term to obtain
\begin{equation}\label{E:scaledEnergyApprox}
\frac{\mathcal{E}[F]}{R^4}\sim\frac{\lambda}{R^2}\left(\frac{n^2}{4\pi}\right)+\frac{2\pi^2}{n}.
\end{equation}
Finally, optimizing with respect to $n$ gives
\begin{equation}\label{E:wrinklesScaling}
n\sim\left(\frac{\lambda}{R^2}\right)^{-1/3},
\end{equation}
which implies that
\begin{equation*}
n\sim R^{2/3}
\end{equation*}
for fixed $\lambda$ and $n\gg1$. Substituting Eq.\eqref{E:wrinklesScaling} into Eq.\eqref{E:scaledEnergyApprox} and solving for $\mathcal{E}$, we obtain
\begin{equation*}
\mathcal{E}\sim R^{10/3},
\end{equation*}
again for fixed $\lambda$ and $n\gg1$.

\begin{figure}[htbp]
\includegraphics[width=0.95\linewidth]{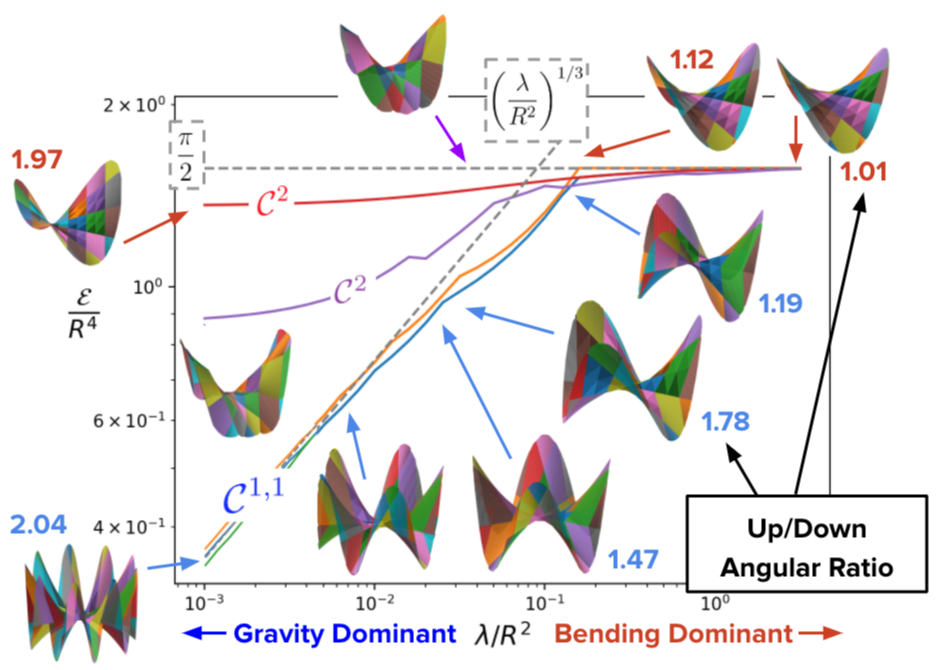}
\caption{Energy scaling of static hyperbolic sheets: There is a significant energy gap between $\mathcal{C}^{1,1}$ and $\mathcal{C}^2$ surfaces in the gravity-dominant regime. Branch points allow for dramatic decreases in gravity energy. The red $\mathcal{C}^2$ curve is the minimal energy for a $\mathcal{C}^2$ surface with straight asymptotic lines. The purple curve is for general $\mathcal{C}^2$ isometries where the asymptotic lines of the out-of-plane displacement surface may be \emph{curved}. The orange, blue, and green curves are for $\mathcal{C}^{1,1}$ isometries. Both the orange and blue curves are for surfaces with a single branch point at the origin with upward- and downward-curving sectors of equal or different angular extent, respectively. The green curve is for surfaces with one generation of offsetted branch points. The number by each image of a hyperbolic sheet is the angular ratio between the upward- and downward-curving sectors}
\label{F:nonDimScaling}
\end{figure}

Fig. \ref{F:nonDimScaling} is a plot of the (scaled) energy as a function of the dimensionless parameter $\lambda/R^2$ for $\mathcal{C}^2$ (no branch point) and $\mathcal{C}^{1,1}$ (with branch points) isometries.  The red curve is for those $\mathcal{C}^2$ isometries with \emph{straight} asymptotic lines. The middle, purple curve is for general $\mathcal{C}^2$ isometries where the asymptotic lines may be \emph{curved}, i.e. the two $u$ (and/or $v$) edges incident on a vertex are not necessarily collinear. The orange and blue curves are for $\mathcal{C}^{1,1}$ isometries with a single branch point at the origin and an out-of-plane displacement surface with straight asymptotic lines. While the orange curve is for $\mathcal{C}^{1,1}$ isometries with upward- and downward-curving sectors restricted to be of equal angular extent, the blue curve is for surfaces that allow these sector angles to be different. The green curve is for $\mathcal{C}^{1,1}$ isometries allowing for one generation of offsetted branch points, i.e. one iteration of surgery as depicted in Fig.~\ref{fig:SelfSimilar}. 

The red and purple curves in Fig. \ref{F:nonDimScaling} correspond, respectively, to a pure quadratic function, with no branch points and identical rhombi, and a piecewise function where all the interior vertices still have degree  four, so there are no branch points, while the rhombi themselves are not necessarily congruent. Unsurprisingly, there is an energy gap between the red and purple curves to the left of the plot (Fig. \ref{F:nonDimScaling}) since the sheet can exploit the additional freedom of varying the shapes of the rhombi to further lower the center of mass thereby decreasing the gravitational potential energy with minimal increases in the bending energy.

The introduction of a single branch point at the origin of increasing degrees, further flattens the sheet by generating an increasing number of wrinkles and, thereby, lowering the gravity energy dramatically. Of course, the accompanying increase in bending energy with higher degree branch points limits how many wrinkles are generated due to a balance of bending and gravity. However, these surfaces are no longer $\mathcal{C}^2$, but rather $\mathcal{C}^{1,1}$.

The energetic difference between the orange and blue curves amounts to only a slight vertical offset with no difference in the scaling $\mathcal{E}\propto\left(\lambda/R^2\right)^{1/3}$. By optimally tuning the angular ratio between the upward- and downward-curving sectors to be closer to 2:1, the surface is able to attain a lower center of mass and, hence, lower the overall energy. Offsetted branch points also allow for decreases in the energy, but numerical experiments suggest that the energy scaling remains the same \cite{Yamamoto2021Role}. 

\subsection{Dynamics of a rolling hyperbolic sheet}\label{sec:rotate}

\begin{figure}[htbp]
   \centering
   \begin{subfigure}[t]{0.48\linewidth}
     \centering
     \includegraphics[trim={2cm, 3.5cm, 2cm, 2.5cm}, clip, width=\textwidth]{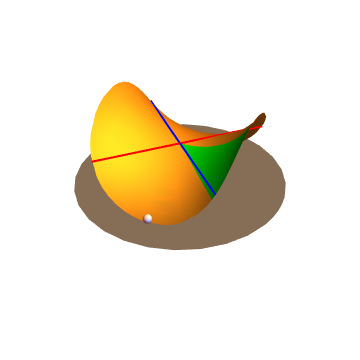}
   \end{subfigure}
   ~
   \begin{subfigure}[t]{0.48\linewidth}
     \centering
     \includegraphics[trim={2cm, 3.5cm, 2cm, 2.5cm}, clip, width=\textwidth]{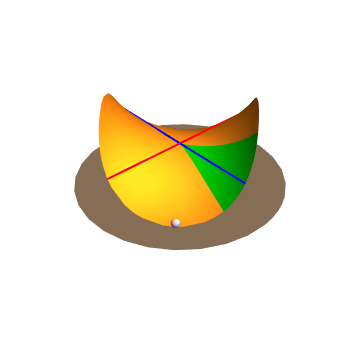}
   \end{subfigure}
   ~
   \begin{subfigure}[t]{0.48\linewidth}
     \centering
     \includegraphics[trim={2cm, 3.5cm, 2cm, 2.5cm}, clip, width=\textwidth]{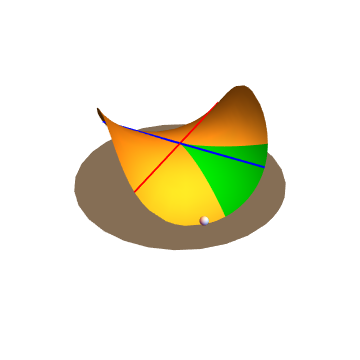}
   \end{subfigure}
   ~
   \begin{subfigure}[t]{0.48\linewidth}
     \centering
     \includegraphics[trim={2cm, 3.5cm, 2cm, 2.5cm}, clip, width=\textwidth]{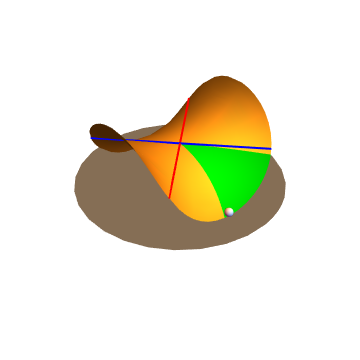}
   \end{subfigure}
   \caption{Dynamics induced by rotating the asymptotic frame relative to the material.}
   \label{fig:3frames}
\end{figure}

The conventional formulation of elasticity models the deformation of an elastic object as a map from the Lagrangian (material) frame $\Omega$ to the Eulerian (lab) frame $\mathcal{V} = \mathbb{R}^3$. Our construction of small-slopes isometries in Section~\ref{sec:small-slope} suggests that there is a third natural frame for thin hyperbolic sheets, namely its asymptotic (skeleton) frame, $\mathcal{A}$. While it is adequate to use conventional formulations of elasticity, it might be profitable to consider formulations that use all three frames. We illustrate this in Fig.~\ref{fig:3frames} with an example which 
shows a smooth saddle, that rolls without slipping. For a hyperelastic material, the deformation costs zero energy and is thus a Goldstone mode although the sheet is not rigid and undergoes an elastic bending deformation. More realistic modeling, following the approach in \cite{acharya2020continuum} will yield a small amount of dissipation per rotation period. In any case, the dynamics is entirely driven by the rotation of the asymptotic frame (the red/blue lines give the ruling) with respect to the material frame (green triangle painted on the material).

Let $(u,v)$ denote the asymptotic coordinates corresponding to two families of rulings whose projections are mutually orthogonal. A solution to $\mathrm{det}(D^2 w) = -1$ with these rulings is given by $x=u,y=v$ and $w=uv$. We can now solve Eq.~\eqref{eq:no_strain} for the in-plane displacements to get
\begin{align*}
\xi & = - \frac{1}{3} u v^2, \qquad  \eta  = - \frac{1}{3} u^2 v.
\end{align*} 
Because we will consider deformations that are expressible as the action of 2d rotations, it is useful to introduce the complex variables $z = u+iv$, $\bar{z} = u-iv$ for the asymptotic coordinates and $\zeta = x+iy$, $\bar{\zeta} = x-iy$ for the material coordinates on the sheet. The Eulerian description of this {\em static} deformation, which includes the in-plane displacements, is given by the mapping $(z,\bar{z}) \mapsto \psi_0(z,\bar{z})$:
\begin{equation*}
\psi_0 = \begin{bmatrix} \zeta + \epsilon^2(\xi+i \eta) \\  \bar{\zeta} + \epsilon^2(\xi -i \eta) \\ w_0 + \epsilon w \end{bmatrix} =  \begin{bmatrix}
z - \dfrac{\epsilon^2}{12}\bar{z}(z^2 - \bar{z}^2) \\[1em] 
\bar{z} - \dfrac{\epsilon^2}{12}z(z^2 - \bar{z}^2) \\[1em] 
w_0 + \dfrac{\epsilon}{4i}(z^2 - \bar{z}^2)
\end{bmatrix},\end{equation*}
where $w_0$ is determined by the condition that the minimum of $w_0 + \dfrac{\epsilon}{4i}(z^2 - \bar{z}^2)$ on the disk $|\zeta| \leq a$, which occurs at $z = \pm a e^{i \pi /4}$, is on the horizontal at height 0, so $w_0 = \frac{a^2}{2}$.

We can now compute the dynamic shape of rotating sheet, as well as its Lagrangian to Eulerian deformation map through
\begin{align}
z & =  e^{i\omega t} \zeta, \quad \bar{z}  = e^{-i\omega t} \bar{\zeta},  \nonumber \\
\psi(t) & =  e^{i A t} \psi_0, \quad A =\begin{bmatrix}
\Omega  & 0 & 0 \\ 0 & - \Omega  & 0 \\ 0 & 0 & 0
\end{bmatrix},
\label{rotate-deform}
\end{align}
where $\Omega$ represents the angular velocity of the ``shape" in the lab frame and $\omega$ represents the angular velocity of the asymptotic frame with respect to the material. The rate at which the material rotates with respect to the lab is therefore $\Omega - \omega$. In Fig.~\ref{fig:3frames}, the angular velocities $\omega$ and $\Omega$ are positive, i.e., the rotation is counterclockwise as viewed from above.

The trajectory of a material particle with Lagrangian coordinates $(\zeta,\bar{\zeta})$ is given by $r(\zeta,\bar{\zeta},t)=e^{i A t} \psi_0(e^{i \omega t} \zeta, e^{-i \omega t} \bar{\zeta})$; therefore, the instantaneous velocity of a material particle is
$$
v(\zeta,\bar{\zeta},t) = i A r(\zeta,\bar{\zeta},t) + e^{i A t} i \omega (\zeta \partial_z \psi_0 - \bar{\zeta}\partial_{\bar{z}} \psi_0).
$$
The rolling condition dictates that points of contact between the rolling body and the surface it contacts must have zero velocity. The points of contact are given by $z = \pm a e^{-i \pi/4}$ and  $\bar{z} = \pm a e^{i \pi/4}$, which correspond to the material points $\zeta = \pm a e^{-i \pi/4 - i \omega t}$ and $\bar{\zeta} =  \pm a e^{i \pi/4 + i \omega t}$. Setting the velocity at these points to 0 we get 
\begin{equation}
 \Omega = \frac{6 + a^2 \epsilon^2}{6 - a^2 \epsilon^2} \, \omega.
\label{frequencies}
\end{equation}
Substituting this relation in Eq.~\eqref{rotate-deform} gives the full description of the deformation of the rotating saddle through a one parameter family of isometries
including the correct imposition of the ``rolling" constraint: an Eulerian condition.

%%%%%%%%%%%%%%%%%%%%%%%%%%%%%%%%%%%%%%%%%%%%%%%%
%
%		Discussion
%
%%%%%%%%%%%%%%%%%%%%%%%%%%%%%%%%%%%%%%%%%%%%%%%%

\section{Discussion} \label{sec:discussion}

In this work we present a general method for constructing exact isometric immersions of hyperbolic metrics with finite bending content in the small-slope setting. We showed that the relevant singularities in free hyperbolic sheets are therefore branch points and lines of inflection. These defects are unique in that the elastic energy does not concentrate on the defects as $t\rightarrow 0$. This is in contrast to crumpled intrinsically flat sheets in which the stretching and bending energies are equipartitioned across across elastic ridges \cite{lobkovsky} and around $d$-cones \cite{benAmar1997Crumpled}. Indeed, the existence of exact small-slope isometries with finite bending content ensure that the elastic energy of global minimizers scales like $t^2$ which is much smaller than $t^{5/3}$, the energy for crumpled intrinsically flat sheets \cite{lobkovsky,conti2008confining}. 

In this work, we develop `geometric' methods that are directly applicable to the non-smooth configurations describing the vanishing thickness, $t \to 0$, limit of hyperbolic sheets. For non-zero but small thickness $t>0$, we expect that the lines of inflection and branch points are smoothed out on a small scale \cite{gemmer2012defects}, and there is an alternative framework for deriving thermodynamically consistent models, including the effects of dissipation, in the $t>0$ setting \cite{acharya2020continuum}.

One consequence of our construction of piecewise smooth isometries is that there are {\em continuous families} of low-energy states obtained by appropriately gluing together isometries and ``floppy modes" that come from variations within these families. Thin hyperbolic sheets are therefore easily deformed by weak stresses and the pattern selected may be sensitive to the dynamics of the swelling process, experimental imperfections, or other external forces. A statistical description of the singularities and their interactions is therefore a natural approach to study this system, in contrast to a static approach that seeks to find the global energy minimum. This problem is indeed very much open; a first step might be to analyze the energy scaling, following the discussion in \S\ref{sec:shapecontrol}, of a surface with a prescribed distribution of branch points. 

The examples we presented in this work are but a small sample of the nontrivial mechanical properties of hyperbolic sheets. This ``extreme mechanics" can be exploited by controlling branch points, which, as we discussed above, are not tied to material points and can thus move ``easily". We conjecture these nonlinear mechanical properties are exploited by a variety of biological systems for locomotion or to effect large morphological changes with a small energy budget. These mechanical properties can therefore confer evolutionary advantages to living organisms and thus contribute to the observed ubiquity of hyperbolic forms in nature \cite{wertheim2016corals}.  

Finally, we believe the results presented in this work extend more generally to hyperbolic Monge-Ampere equations in various geometries, as we have outlined in this paper, but further work is needed for a precise formulation and rigorous proofs of the generalizations. 
\section*{Acknowledgments}
We warmly acknowledge the many insightful conversations we've had with Benny Davidovitch, Eran Sharon, and Ian Tobasco on these topics. We are grateful to Ido Levin who helped make the PVS gel samples and to Eran Sharon for providing the lab space and equipment for the force measurement experiments. KY and SV gratefully acknowledge the hospitality of the Racah Institute of Physics at the Hebrew University, where portions of this work were carried out.  KY also acknowledges funding from the U.S.-Israel Binational Science Foundation Prof. Rahamimoff Travel Grant for the research visit in Israel. TS was partially supported  by a Michael Tabor fellowship from the Graduate Interdisciplinary Program in Applied Mathematics at the University of Arizona.  SV was partially supported by the Simons Foundation through awards 524875 and 560103. KY, ES and SV were partially supported by the NSF award DMR-1923922. KY was also partially supported by the 2018-19 Michael Tabor fellowship from the  Graduate  Interdisciplinary Program in Applied Mathematics at the University of Arizona, the 2020 Marshall Foundation Dissertation Fellowship from the Graduate College at the University of Arizona, and the NSF RTG grant DMS-184026.

\section*{Author Contributions}
KY, TS, JG, and SV developed the theoretical formalism and performed the analytic calculations. KY, TS, and ES performed the numerical
simulations. KY carried out the force measurement experiments and analyzed the data. All the authors discussed the results. All the authors contributed to writing the final version of the manuscript. All of the authors read and approved the final manuscript.
\bibliographystyle{epj}
\bibliography{hyperbolic_sheets.bib}
\end{document}